\documentclass[times,twocolumn,final]{elsarticle}
\usepackage{medima}

\usepackage{soul,color}
\soulregister\citep7
\soulregister\ref7
\soulregister\pageref7

\usepackage{amsfonts}
\usepackage{amsmath}
\usepackage{epsfig}
\usepackage{graphicx}
\usepackage{comment}
\usepackage{amsmath,amssymb} % define this before the line numbering.
\usepackage{color}
\usepackage{subfigure}
\usepackage[]{animate}
\usepackage{animate}
\usepackage{float}
\usepackage{bm}
\usepackage{multirow}
\usepackage{rotating}
\usepackage{wrapfig}
\usepackage{lineno}
\usepackage{epsfig}
\usepackage{nicefrac}       % compact symbols for 1/2, etc.
\usepackage{microtype} 
\usepackage{float}
\usepackage{amsmath}
\usepackage{enumerate}
\usepackage{enumitem} 
\usepackage{amsmath,amsfonts,amssymb}
\usepackage{graphicx}

\usepackage{amsfonts}
\usepackage{graphicx}
\usepackage{amsmath,amssymb}

\usepackage{appendix}

\usepackage[ruled, vlined]{algorithm2e}

\usepackage{lipsum}
\usepackage{array}
\usepackage{amsmath}

\usepackage{amsmath,amssymb,amsfonts}
\usepackage{algorithmic}
\usepackage{graphicx}
\usepackage{textcomp}
\usepackage{hyperref}

\definecolor{newcolor}{rgb}{.8,.349,.1}

\journal{Medical Image Analysis}

\begin{document}

\verso{X. Liu \textit{et~al.}}

\begin{frontmatter}

\title{Attentive Continuous Generative Self-training for Unsupervised Domain Adaptive Medical Image Translation}%

%Adapting an Off-the-Shelf Model for Source-Relaxed Target Image Segmentation

\author[1]{Xiaofeng  \snm{Liu} }

\author[2]{Jerry L. \snm{Prince} }

\author[1]{Fangxu \snm{Xing} }

\author[3]{Jiachen \snm{Zhuo} }

\author[4]{Timothy \snm{Reese} }

\author[3]{Maureen \snm{Stone} }

\author[1]{Georges \snm{El Fakhri}}

\author[1]{Jonghye \snm{Woo}}

\address[1]{Gordon Center for Medical Imaging, Department of Radiology, Massachusetts General Hospital and Harvard Medical School, Boston, MA, 02114}

\address[2]{Department of Electrical and Computer Engineering, Johns Hopkins University, Baltimore, MD, USA} 

\address[3]{Department of Neural and Pain Sciences, University of Maryland School of Dentistry, Baltimore, MD, USA}

\address[4]{Athinoula A. Martinos Center for Biomedical Imaging, Dept. of Radiology, Massachusetts General Hospital and Harvard Medical School, Boston, MA, USA}

\received{20 Nov 2022}
%\received{10 Jan 2023}
\accepted{23 May 2023}
\availableonline{}
\communicated{}

\begin{abstract}
%%%

Self-training is an important class of unsupervised domain adaptation (UDA) approaches that are used to mitigate the problem of domain shift, when applying knowledge learned from a labeled source domain to unlabeled and heterogeneous target domains. While self-training-based UDA has shown considerable promise on discriminative tasks, including classification and segmentation, through reliable pseudo-label filtering based on the maximum softmax probability, there is a paucity of prior work on self-training-based UDA for generative tasks, including image modality translation. To fill this gap, in this work, we seek to develop a generative self-training (GST) framework for domain adaptive image translation with continuous value prediction and regression objectives. Specifically, we quantify both aleatoric and epistemic uncertainties within our GST using variational Bayes learning to measure the reliability of synthesized data. We also introduce a self-attention scheme that de-emphasizes the background region to prevent it from dominating the training process. The adaptation is then carried out by an alternating optimization scheme with target domain supervision that focuses attention on the regions with reliable pseudo-labels. We evaluated our framework on two cross-scanner/center, inter-subject translation tasks, including tagged-to-cine magnetic resonance (MR) image translation and T1-weighted MR-to-fractional anisotropy translation. Extensive validations with unpaired target domain data showed that our GST yielded superior synthesis performance in comparison to adversarial training UDA methods.

%%%%
\end{abstract}

\begin{keyword}
\MSC 41A05\sep 41A10\sep 65D05\sep 65D17
%% Keywords
\KWD Unsupervised Domain Adaptation \sep Deep Self-Training \sep Self-Attention \sep Uncertainty Measurement \sep Cross-modality Translation \sep Medical Image Synthesis.  
\end{keyword}

\end{frontmatter}

\section{Introduction}

\begin{figure}[t]
\begin{center} \textbf{}
\includegraphics[width=1\linewidth]{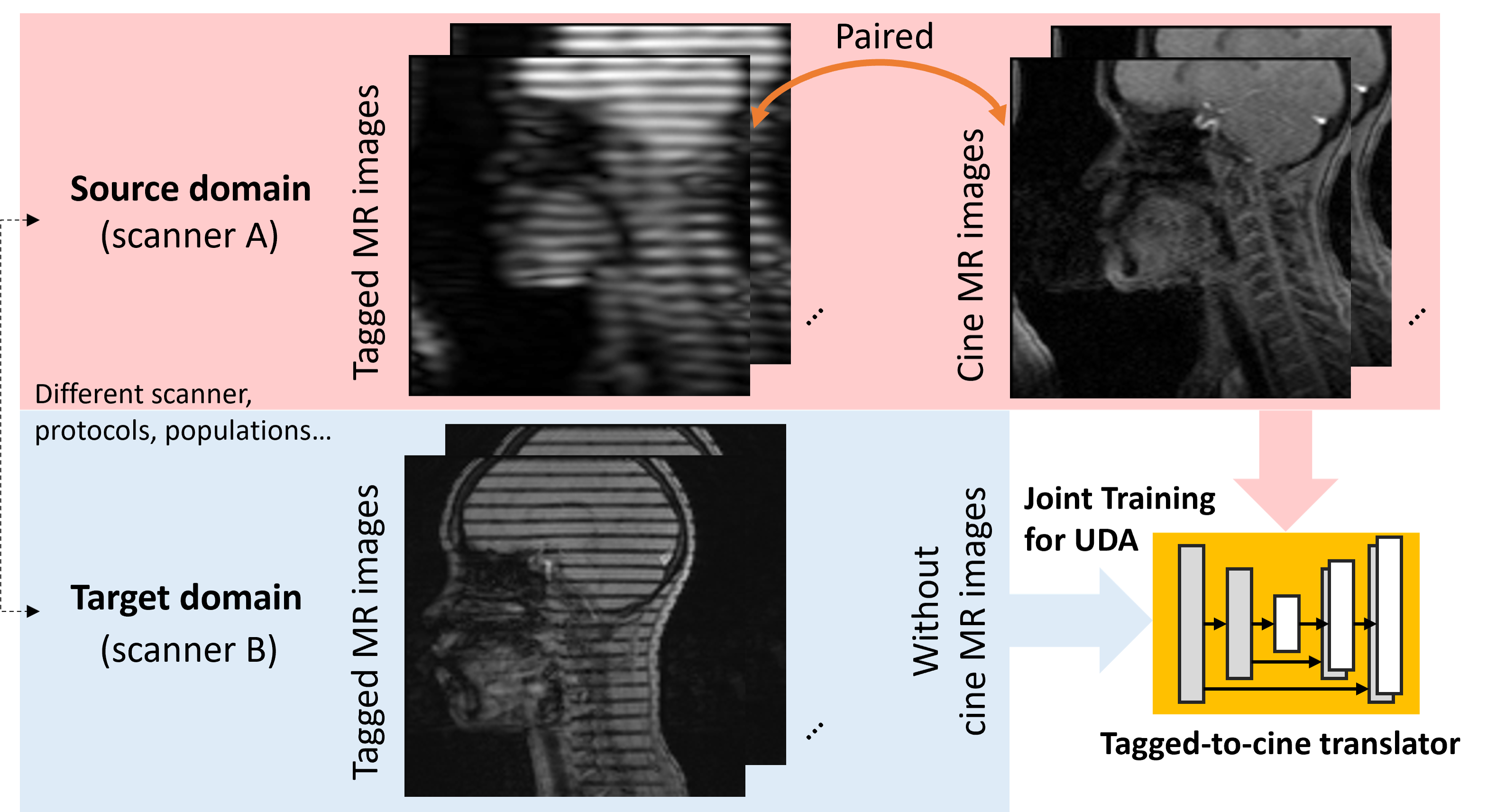}
\end{center}  
\caption{Illustration of a generative UDA task, specifically the cross-scanner tagged-to-cine translation of tongue MR images. The tagged-to-cine image translator is trained jointly with paired source domain and unpaired target domain data to achieve accurate inference in the target domain.}  
\label{intro}\end{figure} 

Image modality translation (also known as image synthesis) has been an active area of research in medical image analysis~\citep{kaji2019overview}. It has been shown to be of great benefit in downstream image analysis tasks, by creating images that either were not acquired due to imaging costs or limitations in study plans or are corrupted by artifacts~\citep{xie2022survey}. Although synthesis performance has improved with advances in deep generative networks~\citep{xie2022survey}, the performance of deep learning models can be degraded if the samples from training and testing datasets are following different distributions~\citep{goodfellow2016deep}. However, the problem of distribution or domain shift~\citep{wang2018deep,xu2021cross} is ubiquitous in medical imaging, since, in many cases, training and testing datasets are acquired using different acquisition protocols or are acquired from different parameters, dose, scanners, or centers. In addition, acquiring sufficient paired data in the new target domain can be expensive and even infeasible~\citep{he2021autoencoder}.  

Therefore, unsupervised domain adaptation (UDA) can be highly desired, which seeks to transfer knowledge from a labeled source domain to an unlabeled target domain that is different but related \citep{wang2018deep}. There is rich literature on UDA across a range of methods and applications~\citep{liu2022deep}. Though UDA has shown great success in discriminative tasks, e.g., classification and segmentation \citep{liu2022deep}, the UDA in image modality translation has not been extensively investigated. The recent work~\citep{he2021autoencoder} proposes to fine-tune a source-domain trained model to an unseen target subject, while the source domain data are not utilized at the UDA stage. To our knowledge, our prior conference version~\citep{liu2021generative} is the first attempt to achieve UDA in medical image modality translation, by jointly optimizing with both paired source and unpaired target domain data. A block diagram of a generative UDA task using cross-scanner tagged-to-cine tongue magnetic resonance (MR) image translation is shown in Fig.~\ref{intro}.

In recent years, self-training has been shown to be a powerful tool for UDA~\citep{zou2019confidence}; in fact, it has outperformed adversarial UDA methods on several discriminative UDA benchmarks, including classification and segmentation as demonstrated in~\cite{wei2021theoretical,mei2020instance,shin2020two,liu2020energy}. Deep discriminative self-training UDA works by first creating a set of one-hot (or smoothed) pseudo-labels in the target domain in an iterative manner. The network is then retrained using these pseudo-labeled target domain samples, as described in~\citep{zou2019confidence}. Self-training UDA for generative tasks, however, has not been extensively studied, and adapting a discriminative self-training UDA model for generative tasks is not a straightforward process. Due to the noisy nature of the outputs of previous iterations in self-training, it is critical to only select the high confidence predictions as reliable pseudo-labels \citep{zou2019confidence}. However, calibrating the uncertainty of generative self-training UDA tasks remains an important yet challenging problem. Defining confidence as the maximum softmax probability is a natural choice in the case of discriminative self-training with a softmax output unit and a cross-entropy loss~\citep{zou2019confidence}. However, the continuous prediction in regression does not inherently possess confidence or reliability confidence or reliability. In addition, in self-training UDA, both {epistemic} and {aleatoric} uncertainties~\citep{kendall2017uncertainties} arise, due to an insufficient amount of data samples in the target domain and unreliable pseudo-labels.

To tackle the above issues, we introduce a generative self-training (GST) framework for UDA medical image synthesis that employs continuous value prediction and regression objectives. A practical variational Bayes learning scheme is proposed to gauge the {epistemic} and {aleatoric} uncertainties in self-training UDA in a unified manner. The supervision of pseudo-labels on target domain data is explicitly regularized by a reliability mask, which is a learnable variable and is determined by the quantified uncertainty. An optimization strategy is then used to alternate between pseudo-label refinement and reliability mask generation, while also retraining the reliability regularized translator network for adaptation. In our prior conference version~\citep{liu2021generative}, we constructed a binary reliability mask, by thresholding the uncertainty map to filter out unreliable pseudo-labels. Although the binary uncertainty selection scheme has been widely used in discriminative self-training and can be adapted to GST~\citep{liu2021generative}, it can lead to sub-optimal performance, due to the following two difficulties: 1)~the relatively rough pseudo-label selection scheme does not fully utilize the well-quantified uncertainty; and 2)~the background region in medical images (e.g., the black boundary in MR images) often has high confidence and dominates the training loss. Of note, the performance on the region of interest is particularly important for medical image synthesis tasks, compared with nature image synthesis tasks.

To address the aforementioned difficulties, in the present work, we build on our prior conference version~\citep{liu2021generative} in the following ways. First, we propose a proper mapping from uncertainty to continuous reliability measurement to achieve fine-grained pseudo-label control and make more efficient use of the well-quantified uncertainty. Second, we incorporate a self-attention scheme into our continuous reliability mask to de-emphasize the background region in our loss calculation.  Third, we present an interpretation of our GST framework in the context of recent works~\citep{he2021autoencoder,zuo2021information,yarram2022joint}. Finally, we extend our evaluation framework beyond the tagged-to-cine MR translation task to include a cross-scanner T1-weighted MRI-to-fractional anisotropy (FA) translation task. To validate and demonstrate the superiority of our proposed GST framework, we compared it with recent UDA methods across all tasks.

The main contributions are summarized as follows:

\noindent$\bullet$ To our knowledge, this is the first attempt to jointly utilize both {paired source domain data and unpaired target domain data} for deep UDA in medical image modality translation scenarios~\citep{liu2021generative}.

\noindent$\bullet$ We propose a novel GST framework which extends the self-training to a generative UDA task, i.e., image modality translation. It is based on adaptive control of the reliability of pseudo-label with a practical Bayesian reliability mask. Furthermore, we systematically study both the aleatoric and epistemic uncertainties in GST UDA.

\noindent$\bullet$ We introduce a self-attention continuous reliability mask that achieves fine-grained pseudo-label control with well-quantified uncertainty, guided by self-learned attention. 

\noindent$\bullet$ Our proposed GST framework is applied to cross-scanner/center and inter-subject tagged-to-cine tongue MR UDA translation, as well as T1-to-FA brain MR UDA translation tasks. These tasks have significant potential to reduce the acquisition time and costs of extra cine/diffusion MR scans. 

Our framework is evaluated using both quantitative and qualitative measures, and the results demonstrate its validity and superiority over conventional adversarial UDA approaches.

\section{Related Work}

\subsection{Medical Image Modality Translation/Synthesis} Image modality translation is a critical pre-processing step in medical image analysis for filling in incomplete modalities or reducing scanning costs~\citep{kaji2019overview}. For example, tagged-to-cine MR image translation has the potential to alleviate the extra cine MR image acquisition time and costs \citep{liu2021dual}. Tagged MR images with horizontal or vertical tag patterns are typically acquired, while the internal organ is in motion, but due to their lower spatial resolution, they may not effectively separate the organ of interest. Therefore, another set of paired cine MR images with higher resolution is usually required as a matching pair, which can double the scanning time and costs~\citep{xing2016analysis}. Multi-modal neuroimaging is another area, where image modality translation is widely used to overcome the missing modality issue caused by motion during the acquisition process \citep{chartsias2017multimodal,gu2019generating,armanious2020medgan,zhou2020hi,zhan2021multi}. Despite the potential applications of medical image translation for UDA in a range of clinical settings, it has not been extensively studied.

\subsection{Cross Domain Medical Image Translation} There are three main categories of related approaches that have been explored to address domain shift in medical image synthesis~\citep{liu2022deep}. The first category involves using labels in both domains within a multi-task learning framework~\citep{zuo2021information}. However, this approach requires the acquisition of sufficient paired data in each new domain, which can be costly. The second category proposes to adapt a pre-trained source model to the target data without access to source data at the adaptation stage~\citep{he2021autoencoder}. The performance of these approaches is generally worse than those with access to source data~\citep{liu2022deep}. The third category is UDA, which jointly trains a model on paired source domain and unlabeled heterogeneous target domain data at the adaptation stage to achieve good testing performance in the target domain~\citep{liu2021generative}. In contrast to the first category that relies on costly paired data acquisition in each new domain, and the second category that may have worse performance without source data at the adaptation stage, UDA fully utilizes both paired source domain and unpaired target domain data. In this work, we focus on UDA for medical image synthesis tasks.

%To our knowledge, our prior conference version~\citep{liu2021generative} is the first attempt to achieve medical image modality translation in UDA scenarios by jointly optimizing with source and target domain data.

%In addition, the challenging test-time UDA is widely used in medical image analysis \citep{karani2021test,he2021autoencoder}, which has only one new collected unpaired target subject in UDA training and testing. The model is expected to achieve good inference on this new subject with the knowledge of different source domain data. Of note, there can be many 2D slices from a subject or its 3D volume. In this work, we focus on test-time UDA to fully utilize the paired source domain and unpaired target domain data. 

\subsection{Unsupervised Domain Adaptation} Early attempts at UDA focused on minimizing a prescribed discrepancy measurement---e.g., maximum mean discrepancy (MMD)~\citep{long2015fully} or batch normalization statistics~\citep{liu2021Off-the-Shelf}---between domains. More recent UDA approaches~\citep{tzeng2017adversarial,liu2021adversarial} incorporated an adversarially learned domain discriminator as a discrepancy measurement. Both early and recent works have primarily focused on discriminative tasks such as classification and segmentation~\citep{wilson2020survey,liu2022deep}. {In medical image segmentation, UDA has achieved a series of superior performances in various applications including vestibular schwannoma and cochlea~\citep{dorent2023crossmoda}, cardiac structures~\citep{li2020dual,zhao2022uda}, and abdominal organs~\citep{zhao2022uda}.}

The recent work~\citep{he2021autoencoder} has shown that the adversarial UDA methods can be adapted to generative tasks, by changing the source domain cross-entropy loss into a regression loss via mean square error (MSE). In addition, adversarial UDA has been used for super-resolution image reconstruction~\citep{wang2021unsupervised}. While aligning latent feature representations can be applied to both discriminative and generative tasks, choosing or learning a proper divergence measure has remained a long-standing problem~\citep{gulrajani2017improved,wilson2020survey}. In addition, adversarial methods tend to be difficult to train and have been shown to sometimes introduce ``hallucination'' content~\citep{goodfellow2016deep}, which is problematic in medical image synthesis applications, where precise and accurate results are crucial.

\subsection{Self-training for UDA} Self-training was originally developed for semi-supervised learning~\citep{zhu2007semi,triguero2015self} and has evolved to include deep embedding learning and classifier adaptation for UDA~\citep{Zou_2018_ECCV}. In recent years, self-training-based UDA has emerged as a powerful method to address unknown labels in the target domain \citep{zou2019confidence}, outperforming adversarial learning-based methods in several discriminative UDA benchmarks~\citep{wei2021theoretical,mei2020instance,shin2020two}. It does not rely on adversarial training, which is difficult to stabilize \citep{goodfellow2016deep}. {Additionally, while self-training UDA has proven effective for classification and segmentation tasks through the use of reliable pseudo-label selection based on the softmax discrete histogram \citep{xie2022unsupervised,cheng2023adpl,kong2022constraining}, the application of the same approach to generative tasks such as image synthesis has not been extensively explored.}

\subsection{Uncertainty Estimation} Quantifying model uncertainty is crucial in medical image analysis~\citep{begoli2019need}. There are two main categories of uncertainty in deep learning models~\citep{kendall2017uncertainties}. First, $aleatoric$ uncertainty accounts for label noise inherent in the dataset and can be modeled by introducing a prior distribution over the model weights to capture the degree of variability of these weights given some data. Second, $epistemic$ uncertainty captures the freedom of the model parameters and can be minimized with sufficient data during training. Bayesian deep learning has advanced uncertainty estimation in recent years~\citep{abdar2021review}, but a systematic analysis of $aleatoric$ and $epistemic$ uncertainties in self-training based UDA settings is lacking. Notably, the uncertainty of the previous round's prediction plays a critical role in measuring the reliability of pseudo-labels in self-training.

\begin{figure*}[t]
\begin{center}
\includegraphics[width=1\linewidth]{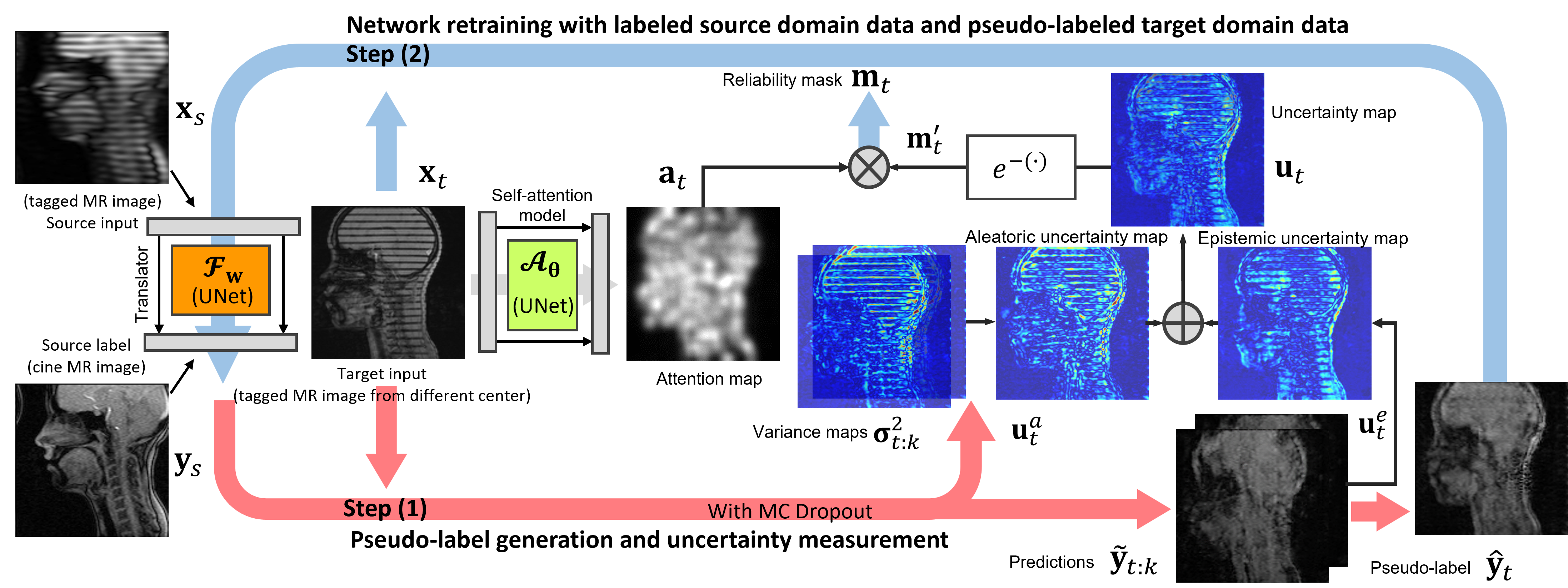}
\end{center}  \vspace{-5pt}
\caption{Illustration of the proposed $\textbf{g}$enerative $\textbf{s}$elf-$\textbf{t}$raining UDA with a self-$\textbf{a}$ttentive $\textbf{c}$ontinuous reliability mask (AC:GST) for tagged-to-cine tongue MR image translation. Our framework involves two steps in each iteration, alternating updates of the networks and the pseudo-labels with their reliability masks.}  
\label{method}\end{figure*} 
 
\section{Methodology} 

In a UDA image translation task, we are given a source domain $\mathcal{D}_S$ and a heterogeneous target domain $\mathcal{D}_T$. From the source domain, we have the paired samples $\{\mathbf{X}_S,\mathbf{Y}_S\}=\{\mathbf{x}_s,\mathbf{y}_s\}_{s=1}^S\sim\mathcal{D}_S$, where $\mathbf{x}_s\in\mathbb{R}^{H\times W}$ and $\mathbf{y}_s\in\mathbb{R}^{H\times W}$, indexed by $s=1, 2,\cdots,S$, are slices with paired/aligned input and output modalities. $H$ and $W$ are the height and width of a slice. For example, in the tagged-to-cine MR translation task, $\mathbf{x}_s$ is a tagged MR slice and $\mathbf{y}_s$ is the corresponding cine MR slice. In addition, $\mathbf{X}_T= \{\mathbf{x}_t\}_{t=1}^T\sim\mathcal{D}_T$ are unpaired target domain samples indexed by $t=1, 2,\cdots, T$. The corresponding target domain labels $\mathbf{Y}_T=\{\mathbf{y}_t\}_{t=1}^T$ are not accessible for training in our UDA task. We aim to learn a translator $\mathcal{F}_{\mathbf{w}}:\mathbf{x}\rightarrow\mathbf{\tilde{y}}$, parameterized by $\mathbf{w}$, for synthesizing an output modality image $\mathbf{\tilde{y}}\in\mathbb{R}^{H\times W}$ that can be generalized well in the target domain. UDA methods are expected to gradually update $\mathcal{F}_{\mathbf{w}}$ for target domain image translation, by utilizing both $\{\mathbf{x}_s,\mathbf{y}_s\}_{s=1}^S$ and $\{\mathbf{x}_t\}_{t=1}^T$.
  
\subsection{Generative Self-training UDA} 

To fill in the missing target domain label $\mathbf{y}_t$, self-training UDA approaches first generate the pseudo-label $\hat{\mathbf{y}}_t\in\mathbb{R}^{H\times W}$ of $\mathbf{x}_t$ in an iterative manner. Then, reliable pseudo-labels are selected as the ``ground-truth" labels for target domain samples, following supervised training protocols. Specifically, traditional self-training approaches typically treat the pseudo-label $\hat{\mathbf{y}}_{t}$ as a learnable latent variable represented by a categorical histogram. This histogram can be either a one-hot distribution~\citep{Zou_2018_ECCV} or a smoothed histogram distribution~\citep{zou2019confidence,liu2020energy}. The pseudo-label is typically derived from the predictions in previous iterations, by selecting the maximum probability category to construct the one-hot label. However, the previous predictions are inherently noisy, which poses a significant challenge for self-training UDA.

The de-facto solution to discriminative self-training is to define a binary selection threshold to progressively select confident pseudo-labels~\citep{zhu2007semi,Zou_2018_ECCV,zou2019confidence}. For classification or segmentation models with categorical softmax output units for an image or pixel, the maximum value of the softmax distribution is commonly used to approximate the confidence of the model's prediction~\citep{zou2019confidence}. The prediction with a maximum softmax output histogram probability higher than a threshold is considered a confident prediction and is used as a pseudo-label for supervised training. On the other hand, if the maximum softmax output histogram probability of a prediction is lower than a threshold, it is considered an uncertain prediction and is set to the all-zero vector $\bm{0}$. Therefore, only the selected confident pseudo-label will contribute to the model update. In the case of continuous output values from a translation model, setting unreliable pseudo-labels to zeros does not remove them from evaluation or prevent their use in the regression loss. This can lead to poor model performance and unreliable synthesis results.

Inspired by discriminative self-training methods~\citep{zou2019confidence,wei2021theoretical,liu2020energy}, we propose a novel GST framework, where the predictions are pixel-wise continuous values and the contributions of pseudo-labels are controlled with a learnable reliability mask $\mathbf{m}_{t}=\{{m}_{t,n}\}_{n=1}^{H\times W}$, where $n$ indexes the pixel in the images. To train this model, the sum of two loss functions is minimized as
\begin{align}\label{111}
&\begin{matrix}\underset{\mathbf{w},\mathbf{m}_t}{\mathop{\min}}\underbrace{{\mathbb{E}}_{\forall s,n\in \mathcal{D}_S} ||y_{s,n}-\tilde{y}_{s,n}||^2_2} + \\ ~~{\mathcal{L}_{reg}^{s}(\mathbf{w})} \end{matrix}\begin{matrix}\underbrace{{\mathbb{E}}_{\forall t,n\in \mathcal{D}_T} ||(\hat{y}_{t,n}-\tilde{y}_{t,n})m_{t,n}||^2_2,}\\ {\mathcal{L}_{reg}^{t}(\mathbf{w},\mathbf{m}_t)}\end{matrix}  \\
&~~\mathrm{s.t.} ~~m_{t,n}\in[0,1],\forall t,n \nonumber
\end{align} where ${{x}}_{s,n},{{y}}_{s,n},{{x}}_{t,n},\hat{{y}}_{t,n}\in [0,255]$ in our implementations. Here, $x_{t,n}$ represents the $n$-th pixel of the $t$-th target domain input image ${\mathbf{x}}_{t}$, and $\tilde{y}_{s,n}$ and $\tilde{y}_{t,n}$ denote the predicted source and target images, respectively. Notably, we have a ground truth label of source domain data in the UDA setting, which has a constant reliability of 1. We indicate the regression losses (in the empirical form of MSE) of the source and target domain data as $\mathcal{L}_{reg}^{s}(\mathbf{w})$ and $\mathcal{L}_{reg}^{t}(\mathbf{w},\mathbf{m}_t)$, respectively. Of note, there is only one translator $\mathcal{F}_\mathbf{w}$, parameterized by $\mathbf{w}$, which is updated with both $\mathcal{L}_{reg}^{s}(\mathbf{w})$ and $\mathcal{L}_{reg}^{t}(\mathbf{w},\mathbf{m}_t)$. 

Instead of setting the uncertain pseudo-labels to zero as in $\hat{y}_{t,n}=\bm{0}$, we use the reliability mask $\mathbf{m}_{t}$ as a learnable variable with respect to uncertainty to control the contribution of pseudo-labels in model training. The pseudo-labels $\hat{y}_{t,n}$ are simply the previous continuous predictions without post-processing. {The core issue in our generative self-training method is to properly associate the continuous prediction uncertainty with $\mathbf{m}_{t}$ to ensure reliable pseudo-label selection.}

\subsection{Bayesian Uncertainty in Self-training UDA} 
%\vspace{-5pt} 

After obtaining the learnable reliability mask $\mathbf{m}_{t}$, the next step is to determine the value of the mask ${{m}}_{t,n}$ for unpaired target sample pixels to reflect the uncertainty of the pseudo-label. While uncertainty estimation has been extensively explored in computer vision and machine learning~\citep{kendall2017uncertainties,carvalho2020scalable,shen2021real}, its application to generative UDA in medical image analysis remains largely unexplored.
 
In learning-based tasks, two types of uncertainty are commonly considered: \textit{aleatoric} uncertainty and \textit{epistemic} uncertainty~\citep{der2009aleatory,kendall2017uncertainties,hu2019supervised}. Aleatoric uncertainty captures the inherent randomness and variability in the observations, while epistemic uncertainty arises from the lack of information or knowledge about the model parameters. In the case of self-training, the generated pseudo-labels are inherently noisy, and therefore characterized by aleatoric uncertainty. Moreover, limited iterations of model training and insufficient training samples can lead to epistemic uncertainty with respect to the model parameters.

In what follows, we take both uncertainties into consideration to provide a holistic uncertainty calibration. {Specifically, epistemic uncertainty is measured via Monte Carlo dropout, and aleatoric uncertainty is included in the network output.} 

We can model the epistemic uncertainty via Bayesian network learning, which hinges on learning a posterior distribution $p(\mathbf{w}|\mathbf{X}_T,\mathbf{\hat{Y}}_T)=\frac{p(\mathbf{\hat{Y}}_T|\mathbf{w},\mathbf{\hat{X}}_T)p(\mathbf{w})}{p(\mathbf{\hat{Y}}_T|\mathbf{\hat{X}}_T)}$ over probabilistic network parameters in place of deterministic network parameters~\citep{rasmussen2003gaussian}. However, we cannot evaluate $p(\mathbf{w}|\mathbf{X}_T,\mathbf{\hat{Y}}_T)$ analytically, since the integration in the normalizing distribution $p({\mathbf{\hat{Y}}_T|\mathbf{\hat{X}}_T})=\int_{\mathbf{w}} p(\mathbf{\hat{Y}}_T|\mathbf{w},\mathbf{\hat{X}}_T) \text{d}\mathbf{w} $ is intractable. One common approach to address this issue is to use a variational approximation to substitute the underlying posterior distribution. Given an approximate distribution $q(\mathbf{w})$ over the parameters of the network, the minimization of the Kullback-Leibler (KL) divergence between $q(\mathbf{w})$ and $p(\mathbf{w}|\mathbf{X}_T,\mathbf{\hat{Y}}_T)$, i.e., $\textrm{KL}(q(\mathbf{w})||p(\mathbf{w}|\mathbf{X}_T,\mathbf{\hat{Y}}_T))$, can be enforced to learn the approximation. In practice, to approximate $q(\mathbf{w})$, dropout variational inference can be applied, by adopting a Bernoulli distribution~\citep{gal2015bayesian}; for example, Monte Carlo (MC) dropout can be used by $K$ times predictions with independent dropout sampling. In this work, we use the MSE to measure the epistemic uncertainty as in~\citep{rasmussen2003gaussian}, which assesses a one-dimensional regression model similar to~\cite{fruehwirt2018bayesian}. Then, $K$ times MC dropout predictions are used to estimate the epistemic uncertainty as 
\begin{align}\label{222} 
u^{e}_{t,n} = \frac{1}{K} \sum_{k=1}^{K} || {\tilde{y}_{t,n:k}-\mu_{t,n}}||^2_2, 
\end{align} 
where $k\in\{1,\cdots,K\}$ indexes the MC dropouts, and $\mu_{t,n}=\frac{1}{K} \sum_{k=1}^{K} \tilde{y}_{t,n:k}$ is the predictive mean of all MC dropout predictions $\{\tilde{y}_{t,n:k}\}_{k=1}^K$. 

In addition, it is necessary to model the heteroscedastic aleatoric uncertainty for each sample $\mathbf{x}_t$~\citep{nix1994estimating,le2005heteroscedastic}. To this end, we split the decoder part of $\mathbf{w}$ to predict both $\tilde{\mathbf{y}}_t$ and a variance map $\bm{\sigma}^2_t\in\mathbb{R}^{H\times W}$ which has element $\sigma^2_{t,n}$ as the predicted variance of the $n$-th pixel~\citep{le2005heteroscedastic,kendall2017uncertainties}. Notably, the ``aleatoric uncertainty labels" are not available for supervised training. Instead, we are able to learn the variance map $\bm{\sigma}^2_t$ implicitly from a regularized regression loss function~\citep{le2005heteroscedastic,kendall2017uncertainties}. Specifically, the target domain loss can be formulated as 
\begin{align}\label{333}\mathcal{L}_{reg}^{t}(\mathbf{w},\mathbf{m}_t,\sigma^2_{t})=\frac{1}{T\times N} {\sum\limits_{t=1}^{T}} {\sum\limits_{n=1}^{N}}  (&\frac{1}{\sigma^2_{t,n}}||(\hat{y}_{t,n}-\tilde{y}_{t,n})m_{t,n}||^2_2 \nonumber\\& + \beta\text{log} \sigma^2_{t,n}),\end{align} where the first term is the variance normalized MSE loss with reliability mask control, and the second term is an uncertainty regularization term, $\beta\text{log} \sigma^2_{t,n}$, which prevents the network from predicting infinite uncertainty for any given sample.~ Following~\cite{le2005heteroscedastic,kendall2017uncertainties}, the averaged aleatoric uncertainty of $K$ MC dropout predictions is quantified as  
\begin{align} u^{a}_{t,n}=\frac{1}{K} \sum_{k=1}^{K}\sigma^2_{t,n:k}.\end{align} 
Since the loss in Eq.~(\ref{333}) is to be minimized, it can be 
interpreted as the Lagrangian (where $\beta$ is a Lagrange multiplier) of the following constrained minimization problem:  
\begin{align} &\underset{\mathbf{w}}{\mathop{\min }}\frac{1}{T\times N}{\sum\limits_{t=1}^{T}} {\sum\limits_{n=1}^{N}}  \frac{1}{\sigma^2_{t,n}}||(\hat{y}_{t,n}-\tilde{y}_{t,n})m_{t,n}||^2_2,\\ &\mathrm{s.t.}~\frac{1}{T\times N}{\sum\limits_{t=1}^{T}} {\sum\limits_{n=1}^{N}} \text{log} \sigma^2_{t,n}< \tau, \end{align} 
where $\tau\in\mathbb{R}^+$ indicates the strength of the applied constraint\footnote{It can be rewritten as $\underset{\mathbf{w}}{\mathop{\min}}~ \mathcal{L}=\{\frac{1}{T\times N}{\sum\limits_{t=1}^{T}} {\sum\limits_{n=1}^{N}}  \frac{1}{\sigma^2_{t,n}}||(\hat{y}_{t,n}-\tilde{y}_{t,n})m_{t,n}||^2_2+\beta(\frac{1}{T\times N}{\sum\limits_{t=1}^{T}} {\sum\limits_{n=1}^{N}} \text{log} \sigma^2_{t,n}-\tau)\}$. Since $\beta,\tau\geq0$, an upper bound on $\mathcal{L}$ can be obtained as $\mathcal{L}\leq \mathcal{L}_{reg}^t$.}. Then, the term ${\sum\limits_{n=1}^{N}}\text{log} \sigma^2_{t,n}$ can be interpreted as the measurement of image-level aleatoric uncertainty. As such, the constraint term in Eq. (6) can effectively constrain the target domain's predictive aleatoric uncertainty, which is particularly useful for adaptation training~\citep{han2019unsupervised}. Taken together, the pixel-wise uncertainty estimated in our GST UDA can be formulated as a combination of the two uncertainties~\citep{kendall2017uncertainties}:  
\begin{align} u_{t,n}=u^{e}_{t,n}+u^{a}_{t,n}.\end{align} 

\subsection{Binary Reliability Mask for GST} 

It is possible to use uncertainty measurements in discriminative self-training methods, by either keeping or filtering out the confident or uncertain pseudo-labels, respectively~\citep{triguero2015self,zou2019confidence}. Specifically, these methods propose to construct a binary reliability mask as shown in Eq. (\ref{111}) by:   
\begin{align}\label{bm}
m_{t,n}=\begin{cases}1 & u_{t,n} < \epsilon\\0 & u_{t,n} > \epsilon \end{cases}; ~~\epsilon=\text{min}\{\text{top}~\rho\%~\text{sorted}~u_{t,n}\},
\end{align} where the value of the reliability mask ${{m}}_{t,n}$ is selected by the measured uncertainty ${{u}}_{t,n}$ with a critical threshold $\epsilon$. Since the pseudo-labels are gradually refined along with the adaptation training, ideally $\epsilon$ should be adaptively updated to control pseudo-label learning and selection. Therefore, $\epsilon$ is typically defined by a meta portion parameter $\rho\in[0,1]$, indicating the portion of reliable pseudo-labels. Then, in each iteration, we determine $\epsilon$ by sorting ${{u}}_{t,n}$ in increasing order so that $\epsilon$ is the minimum ${{u}}_{t,n}$ of the top $\rho$ percentile rank~\citep{zou2019confidence,liu2020energy}. This is akin to self-paced learning \citep{kumar2010self,tang2012shifting,zou2019confidence}, where weights are learned following an easy-to-hard strategy. 

With the binary reliability mask definition above, we can now turn to the task of optimizing our self-training objective in 
Eq.~(\ref{111}). Since direct optimization of the self-training objective in Eq.(\ref{111}) is difficult, we leverage the deterministic annealing expectation maximization (EM) algorithm defined in~\cite{grandvalet2006entropy} instead. Specifically, we can solve our GST UDA framework using an alternating optimization scheme consisting of the following two steps:

\vspace{+5pt}
\noindent \textbf{1) Pseudo-label and reliability mask generation.} \label{a)}With the current $\mathcal{F}_{\mathbf{w}}$, apply MC dropout $K$ times for each $\mathbf{x}_{t}$ and measure $u_{t,n}$. Then, calculate the binary reliability mask ${\mathbf{m}}_t$, given the current threshold $\epsilon$. The pseudo-label of the selected pixel in this round is set as $\hat{{y}}_{t,n}={{\mu}}_{t,n}$, which is the average value of $K$ MC dropout predictions.

\vspace{+5pt}
\noindent \textbf{2) Translator $\mathcal{F}_{\mathbf{w}}$ retraining}.\label{b)}~Fix $\hat{\mathbf{Y}}_T=\{\hat{\mathbf{y}}_t\}_{t=1}^{T}$, ${\mathbf{M}}_T=\{{\mathbf{m}}_t\}_{t=1}^{T}$ and update $\mathbf{w}$ by solving:
\begin{align}\label{st_b}
\underset{\mathbf{w}}{\mathop{\min }}&~~ \mathcal{L}_{GST} = \frac{1}{S\times N}{\sum\limits_{t=1}^{S}}{\sum\limits_{n=1}^{N}}  ||y_{s,n}-\tilde{y}_{s,n}||^2_2   \\&+\frac{1}{T\times N}{\sum\limits_{t=1}^{T}}{\sum\limits_{n=1}^{N}}  (\frac{1}{\sigma^2_{t,n}}||(\hat{y}_{t,n}-\tilde{y}_{t,n})m_{t,n}||^2_2+ \beta\text{log} \sigma^2_{t,n}). \nonumber
\end{align} 

We define a round of self-training as the sequential execution of steps \textbf{1)} and \textbf{2)}. Intuitively, step \textbf{1)} is equivalent to performing simultaneous pseudo-label generation and reliability-based selection. To solve step \textbf{2)}, we use a conventional gradient descent algorithm. Additionally, we linearly increase the meta portion parameter $\rho$ from 30\% to 80\% during training to gradually select more pseudo-labels, following the approach proposed in~\cite{zou2019confidence,liu2020energy}.

\subsection{Continuous Reliability Mask}  
  
In the previous section, we described the binary uncertainty selection scheme, which is widely used in discriminative self-training~\citep{zou2019confidence,wei2021theoretical,liu2020energy}. The primary reason for choosing a binary pseudo-label in discriminative self-training is that the softmax prediction itself is not a good measure of uncertainty or confidence~\citep{goodfellow2016deep}, which makes it unreliable to use for controlling uncertainty~\citep{zou2019confidence, liu2020energy}. However, when a well-quantified uncertainty measure, such as our $u_{t,n}$, is available, we should not have to binarize it for use in the framework described above. The challenge, however, is how to use it in a continuous manner.

\begin{center}
\begin{algorithm}[t]
\SetAlgoLined
\scriptsize 
\caption{The training protocol for AC:CST.}
\SetKwInOut{Input}{Input}
\SetKwInOut{Output}{Output}
\textbf{Input:} $\{\mathbf{X}_S, \mathbf{Y}_S\}$, $\{\mathbf{X}_T\}$, $\beta$, $K$, and initialized $\mathcal{F}_{\mathbf{w}}$, $\mathcal{A}_{\theta}$;\\

Source domain supervised pre-training with $\{\mathbf{X}_S, \mathbf{Y}_S\}$ for $\mathcal{F}_{\mathbf{w}}$.

\For{$i\leftarrow m$ \KwTo $M$ \text{iteration in UDA stage}}{

\textbf{{Sample}} a batch of images $\{\mathbf{x}_s,\mathbf{y}_s\}_{s=1}^S$, $\{\mathbf{x}_t\}_{t=1}^T$;\\ 

\For{Step (1)}{
\textbf{{$\mathcal{F}_{\mathbf{w}}$}} with MC dropout: $\mathbf{x}_t\rightarrow\{\bm{\tilde{y}}_{t:k}\}_{k=1}^K$, $\{\bm{\sigma}^2_{t:k}\}_{k=1}^K$;\\
$u^{e}_{t,n} = \frac{1}{K} \sum_{k=1}^{K} || {\tilde{y}_{t,n:k}-\mu_{t,n}}||^2_2$;\\ $u^{a}_{t,n}=\frac{1}{K} \sum_{k=1}^{K}\sigma^2_{t,n:k}$;\\
$u_{t,n}=u^{e}_{t,n}+u^{a}_{t,n}$, $m'_{t,n} = e^{-u_{t,n}};$\\
\textbf{{$\mathcal{A}_{\theta}$}} predict attention map: $\mathbf{x}_t\rightarrow\bm{a}_t = \{a_{t,n}\}_{n=1}^N$;\\
$m_{t,n}=a_{t,n}\cdot m'_{t,n}, \forall t,n$ for self-attention reliability mask $\bm{m}_t$;\\
\textbf{Update} pseudo label $\bm{\hat{y}}_{t}$ with  $\hat{y}_{t,n}=\mu_{t,n}=\frac{1}{K} \sum_{k=1}^{K} \tilde{y}_{t,n:k}$.\\
}

\For{Step (2)}{
\textbf{{$\mathcal{F}_{\mathbf{w}}$}} without MC dropout: $\mathbf{x}_s\rightarrow\bm{\tilde{y}}_{s}$, $\mathbf{x}_t\rightarrow\bm{\tilde{y}}_{t}, \bm{\sigma}^2_{t}$;\\
\textbf{Update} network parameters $\mathcal{F}_{\mathbf{w}}$ with $\mathcal{L}_{GST}$ in Eq.~(9).\\
\textbf{Update} network parameters $\mathcal{A}_{\theta}$ with $\mathcal{L}_{GST}$ in Eq.~(9).\\
}
}
\textbf{Testing:} only apply the trained $\mathcal{F}_{\mathbf{w}}$ without MC dropout.
\end{algorithm}\label{alg_code}
\end{center} 
\vspace{-30pt}

%While the binary uncertainty selection scheme has been widely used in discriminative self-training~\citep{zou2019confidence,wei2021theoretical,liu2020energy} and can be adapted to the GST~\citep{liu2021generative}, it suffers from the following two difficulties: 1) the well-quantified uncertainty is not seamlessly incorporated into the relatively rough pseudo-label selection process; and 2) the background region usually has high confidence and is therefore likely to dominate the training loss. The reasons for these difficulties are first described below, and then our approach to addressing them is explained. 

We propose constructing a continuous reliability mask $m_{t,n}$ for the pseudo label $\hat{y}_{t,n}$, by appropriately mapping the value of $u_{t,n}\in[0,+\infty)$. Although the reciprocal term $1/u_{t,n}$ is a possible candidate for mapping $u_{t,n}\in[0,+\infty)$ to $m_{t,n}$, it can be numerically unstable for $u_{t,n}=0$. Moreover, the background region in many medical images often contains numerous pixels with zero $u_{t,n}$, making this approach unsuitable. We note that translation of the background region is straightforward via an identical mapping, and the network is always highly confident about its prediction in this region. Based on the observations above, we propose a practical solution with a self-attention guided continuous reliability mask. In particular, we propose to add a heuristic non-linear normalization unit to the calculated element in the uncertainty map $u_{t,n}\in[0,+\infty)$ to generate the element of the reliability mask:
\begin{align}
    m'_{t,n} = e^{-u_{t,n}}, ~m'_{t,n}\in(0,1].
\end{align}
With the continuous reliability mask, we are able to achieve fine-grained control over the binary setting to efficiently utilize the well-quantified uncertainty. Specifically, we can view the binary mask as a simplified special case of the continuous mask obtained by applying a threshold.

\subsection{Self-attentive Continuous GST}

%~\citep{liu2019hard}
In addition to our use of a continuous mask, we propose to emphasize anatomical structures rather than the background region in MR data. Notably, the zero or small $u_{t,n}$ values in the background region correspond to the high reliability weights, and are therefore likely to dominate the training loss $\mathcal{L}_{reg}^{t}(\mathbf{w},\mathbf{m}_t)$. In practice, the confidence of pixels in the background region is usually top-ranked to have $m_{t,n}=1$, when using either binary reliability mask in Eq.~(\ref{bm}) or continuous reliability mask $m'_{t,n} = e^{-u_{t,n}}$. These easy samples may dominate the training of the deep network. Therefore, we introduce a self-trained attention network $\mathcal{A}_{\bm{\theta}}:\mathbf{x}_t\rightarrow\bm{a}_t = \{a_{t,n}\}_{n=1}^N$ such that each slice is processed with $\mathcal{A}_{\bm{\theta}}$ to yield the refined corresponding attention map $\bm{a}_t\in\mathbb{R}^{H\times W}$, where $a_{t,n}\in[0,1], \forall t,n$. The attention map $\bm{a}_t$ is then multiplied by the generated reliability mask $m'_{t,n}$ to obtain the attention reliability mask

\begin{align}
    m_{t,n}=a_{t,n}\cdot m'_{t,n}, ~  m_{t,n}\in[0,1].
\end{align}
In this way, the background region is not emphasized during training. We adopt a conventional 2D encoder-decoder structure for $\mathcal{A}_{\bm{\theta}}$, and it is jointly optimized with $\mathcal{F}_\mathbf{w}$ in a collaborative manner, by minimizing the same loss. As a result, our proposed method encourages $\mathcal{A}_{\bm{\theta}}$ to adaptively learn and retain the essential information required for optimal translation performance in a self-training manner. Importantly, our approach does not require an additional attention label for training $\mathcal{A}_{\bm{\theta}}$. The two-step optimization of a continuous reliability mask with an adaptively learned attention module can be formulated as follows:

\vspace{+5pt} 
\noindent \textbf{1) Pseudo-label and attentive reliability mask generation.} \label{a)}With the current $\mathcal{F}_{\mathbf{w}}$, we apply MC dropout $K$ times for each $\mathbf{x}_{t}$ to measure $u_{t,n}$. We then use $\mathcal{A}_{\bm{\theta}}$ to generate the attention map $\bm{a}_t$, and use a heuristically non-linear normalization unit to calculate the continuous reliability mask ${\mathbf{m}}_t$. We do not need to set a threshold $\epsilon$ or a proportion parameter $\rho$ for pseudo-label selection. The pseudo-label of the selected pixel in this round is set to $\hat{{y}}_{t,n}={{\mu}}_{t,n}$, which is the average value of $K$ MC dropout predictions.

\vspace{+5pt} 
\noindent \textbf{2) Translator $\mathcal{F}_{\mathbf{w}}$ and attention module $\mathcal{A}_{\bm{\theta}}$ retraining}.\label{b)}~Fix $\hat{\mathbf{Y}}_T=\{\hat{\mathbf{y}}_t\}_{t=1}^{T}$, ${\mathbf{M}}_T=\{{\mathbf{m}}_t\}_{t=1}^{T}$ and update $\mathbf{w},{\bm{\theta}}$ by solving:
\begin{align}\label{st_b}
\underset{\mathbf{w},{\bm{\theta}}}{\mathop{\min }}&~~ \mathcal{L}_{GST} = \frac{1}{S\times N}{\sum\limits_{t=1}^{S}}{\sum\limits_{n=1}^{N}}  ||y_{s,n}-\tilde{y}_{s,n}||^2_2   \\&+\frac{1}{T\times N}{\sum\limits_{t=1}^{T}}{\sum\limits_{n=1}^{N}}  (\frac{1}{\sigma^2_{t,n}}||(\hat{y}_{t,n}-\tilde{y}_{t,n})m_{t,n}||^2_2+ \beta\text{log} \sigma^2_{t,n}). \nonumber
\end{align}   

The overall training protocol is outlined in Algorithm 1. During testing, only the trained translator $\mathcal{F}_\mathbf{w}$ is used for inference. The continuous reliability measurement not only provides fine-grained control, but also allows us to integrate the attention module to de-emphasize the background regions.

\section{Experiments and Results} 

To demonstrate the effectiveness of our framework, we evaluated it on two translation tasks, including cross-scanner/center tagged-to-cine tongue MR image UDA translation and T1-to-FA brain MR image UDA translation. We followed the test time UDA setting as in~\cite{karani2021test,he2021autoencoder,liu2021generative}, which only uses one unpaired target subject in UDA training and testing. Our framework was implemented on an NVIDIA V100 GPU using the PyTorch deep learning toolbox. For GST training, we used the Adam optimizer with a momentum of 0.5 and a fixed learning rate of $1 \times 10^{-3}$ throughout all experiments. In each iteration, the batch size was set to 16 slices in both the source and target domains. 
 
\begin{figure}[t]
\begin{center}
\includegraphics[width=1\linewidth]{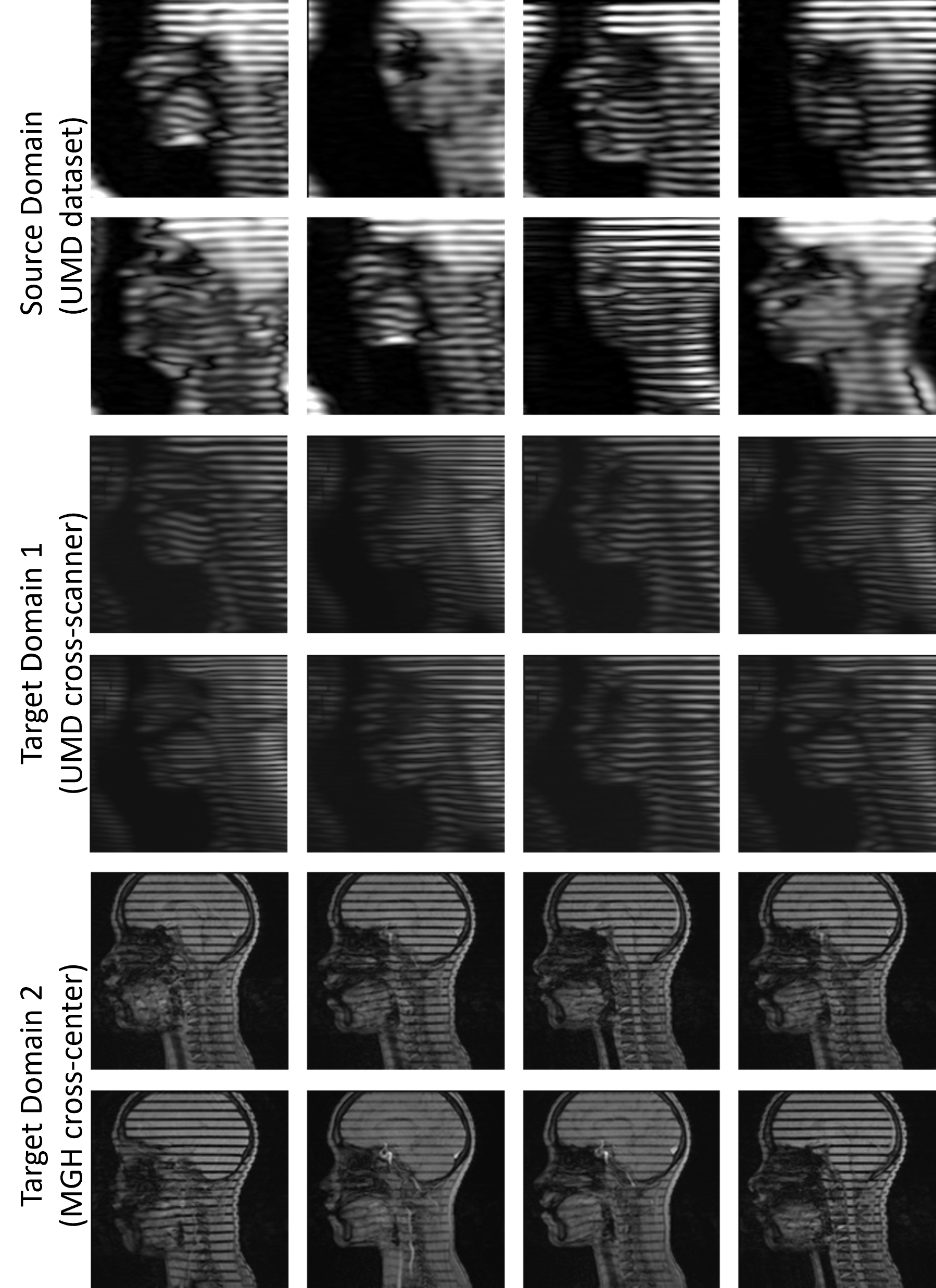}
\end{center}
\caption{Comparison of the collected tagged MR images in different domains.} 
\label{fig:sup1}
\end{figure}

\begin{figure*}[t]
\begin{center}  
\includegraphics[width=1\linewidth]{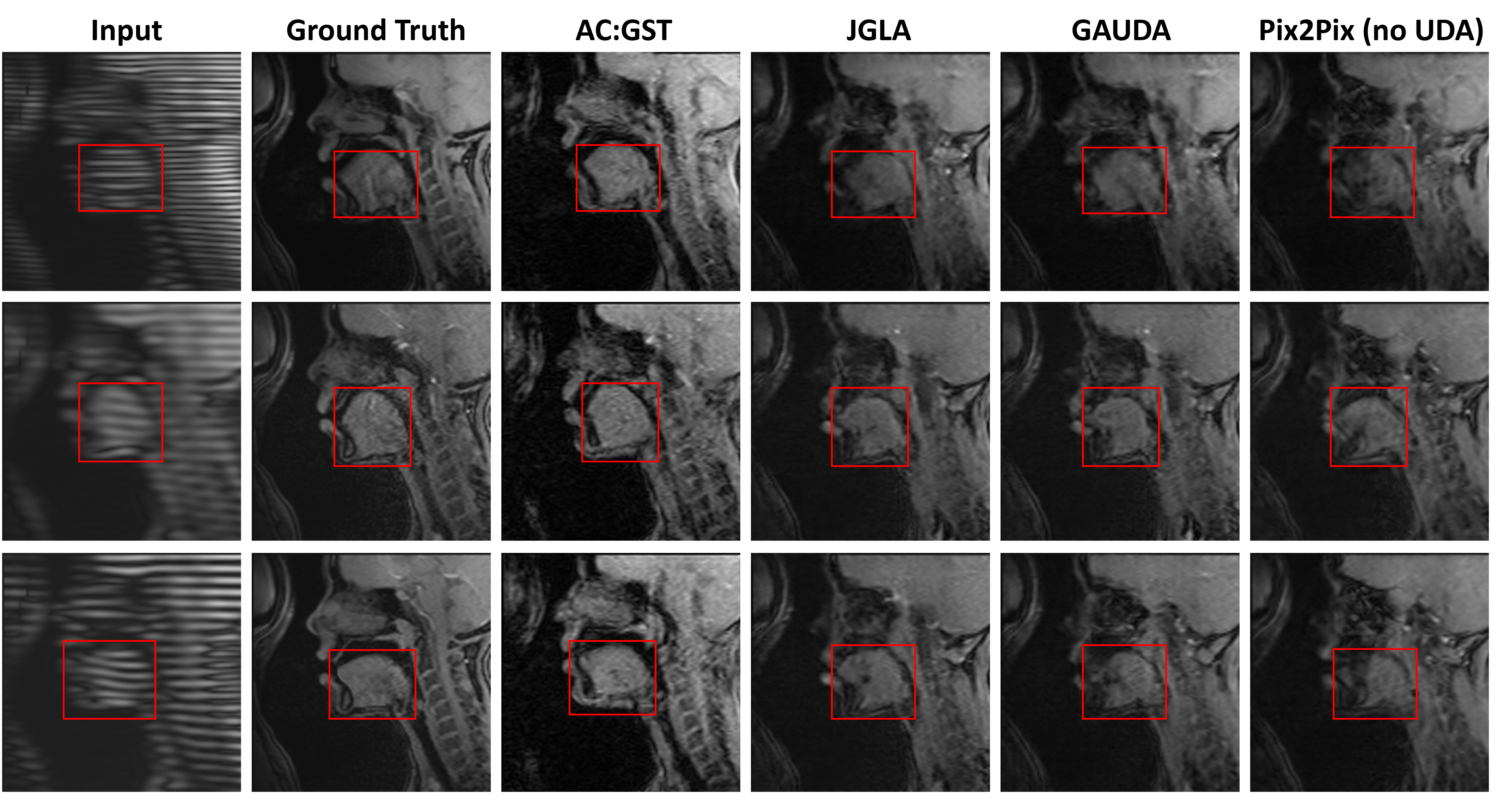} 
\end{center}  
\caption{Comparisons with the adversarial UDA~\citep{cui2020gradually,yarram2022joint} and Pix2Pix~\citep{isola2017image} w/o adaptation for the within UMB cross-scanner tongue tagged-to-cine MR image translation. The red rectangle indicates the tongue region used for tongue motion analysis.} 
\label{fig:results1}
\end{figure*}     
    
\begin{figure*}[t]
\begin{center}  
\includegraphics[width=1\linewidth]{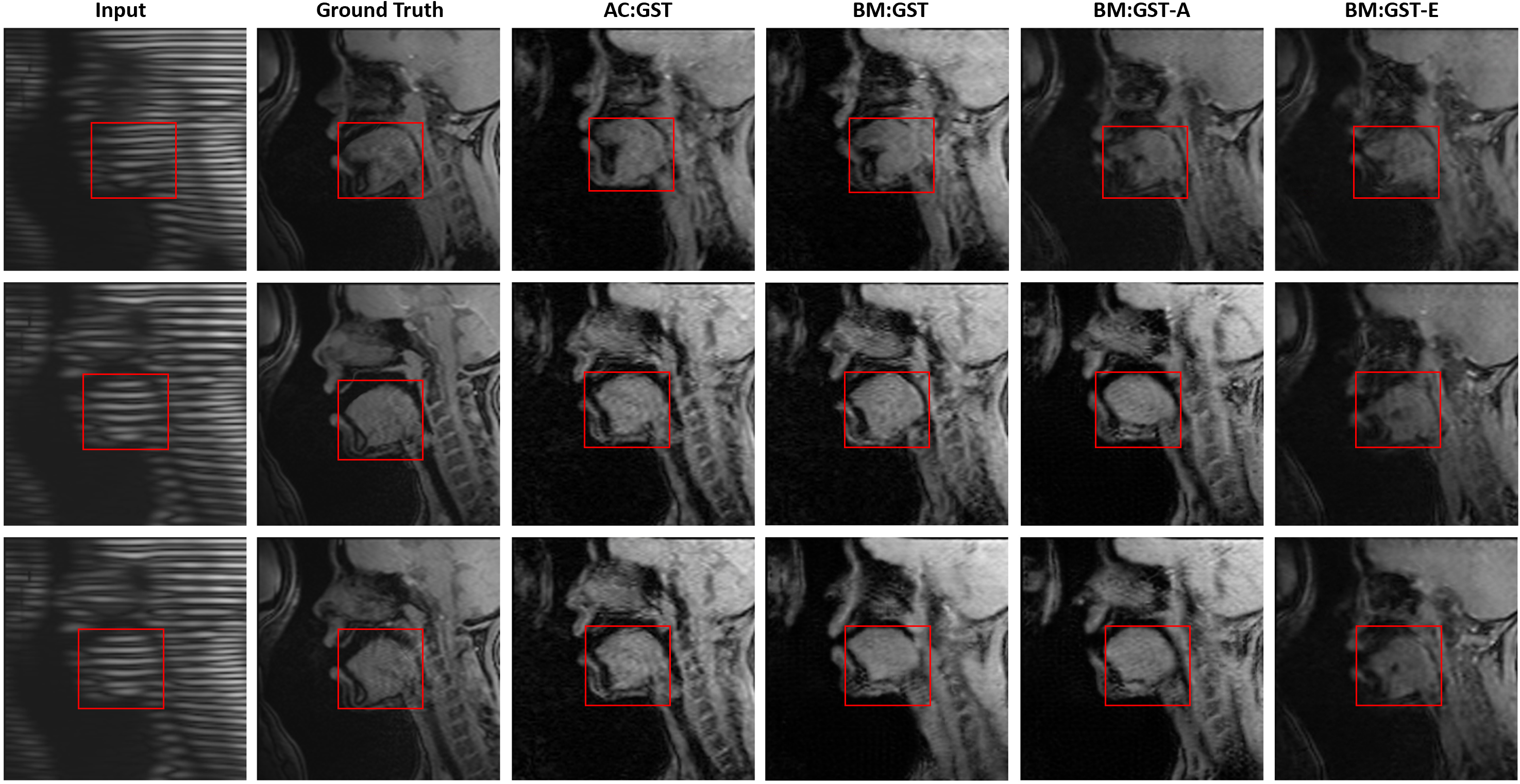} 
\end{center}  
\caption{Ablation studies of our proposed AC:GST, BM:GST, BM:GST-A, and BM:GST-E for within UMB cross-scanner tagged-to-cine tongue MR image translation. The red rectangle indicates the tongue region used for tongue motion analysis.} 
\label{fig:results2}
\end{figure*} 

We denote our proposed framework as AC:GST, which uses the self-attention continuous reliability mask, and BM:GST, which uses the binary reliability mask as in~\citep{liu2021generative}. For our ablation studies, we evaluated AC:GST without the attention scheme (denoted as AC:GST-$\mathcal{A}_{\bm{\theta}}$) and without the continuous reliability mask (denoted as AC:GST-C). Additionally, we carried out experiments, where the epistemic or aleatoric uncertainty is ignored in our uncertainty quantification, which are denoted as -E or -A, respectively. We report the results as mean$\pm$SD over three evaluations.

Similar to~\cite{he2021autoencoder,liu2021generative}, we compared our framework with several widely used discriminative adversarial UDA methods, including ADDA~\citep{tzeng2017adversarial}, GAUDA~\citep{cui2020gradually}, and JGLA~\citep{yarram2022joint}. To adapt these methods to our task, we replaced their output unit and used the MSE loss. Additionally, we evaluated our framework against ASS~\citep{he2021autoencoder}, which fine-tunes a source-domain trained model to an unseen target subject without utilizing the source domain data during the adaptation stage. 

For a fair comparison, we used the UNet-based translator in Pix2Pix~\citep{isola2017image} as the backbone of our GST and the compared methods. We initialized $\mathcal{F}_{\bm{w}}$ with pre-training in the source domain~\citep{liu2021dual}. Note that the last three layers in the decoder were duplicated for variance prediction. We trained $\mathcal{A}_{\bm{\theta}}$ from scratch with random initialization, using the same backbone as $\mathcal{F}_{\bm{w}}$.

\subsection{UMB cross-scanner tagged-to-cine MR translation}
 
\begin{figure}[t]
\begin{center}  
\includegraphics[width=1\linewidth]{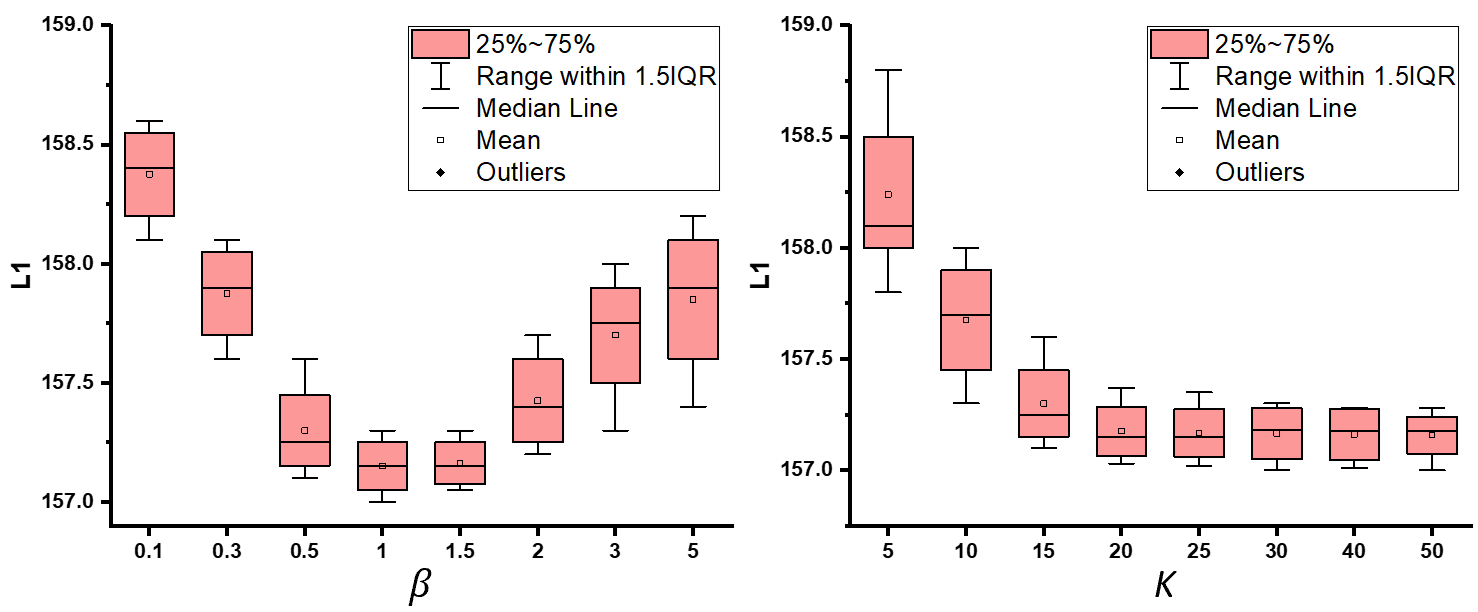}
\end{center}  \vspace{-10pt}
\caption{Sensitivity analysis of $\beta$ and $K$ in AC:GST with respect to L1 error for UMB cross-scanner tagged-to-cine MR translation.}  
\label{sensi1}\end{figure}

Tagged MR images have been widely used to quantify tissue deformation in moving organs, including the heart~\citep{osman2000imaging} and the tongue~\citep{xing20133d,parthasarathy2007measuring}. In this work, we applied our framework to tongue motion data acquired during speech. While internal motion of the tongue can be reconstructed from these tagged images, it is difficult to reconstruct the tongue surface due to the relatively low resolution of the images. Therefore, a set of cine MR images is typically acquired in the same scanning session to provide higher resolution images of the tongue surface during the same speech phrases. If tagged-to-cine MR translation could be performed, then it would not be necessary to acquire the cine MR images, reducing extra acquisition time and ensuring accurate registration between the two sets of image sequences~\citep{liu2021dual}. Fig. \ref{fig:sup1} shows a visual illustration of the appearance difference. The target domain 1 refers to tagged MR images collected from the University of Maryland, Baltimore (UMB) with a different scanner, while the target domain 2 refers to tagged MR images collected from MGH with a different scanner. We used tagged MR images with horizontal tag patterns for our evaluation.

\begin{table}[t]
\centering
\caption{Numerical comparisons and ablation studies of the UMB cross-scanner tongue tagged-to-cine MR translation task. } \vspace{+3pt}
\resizebox{1\linewidth}{!}{
\begin{tabular}{l|c|c|c|c}
\hline

Methods& L1~$\downarrow$& SSIM~$\uparrow$ &	PSNR~$\uparrow$ &  IS~$\uparrow$\\\hline\hline
w/o UDA \citep{isola2017image}     	& {176.4}$\pm$0.1&	0.8325$\pm$0.0012&	26.31$\pm$0.05&	8.73$\pm$0.12  \\\hline

ASS \citep{he2021autoencoder} & {170.5}$\pm$0.3&	0.8525$\pm$0.0011& 30.12$\pm$0.06&	9.95$\pm$0.13  \\\hline

ADDA \citep{tzeng2017adversarial}	&168.2$\pm$0.2&	0.8784$\pm$0.0013&	33.15$\pm$0.04&	10.38$\pm$0.11\\
GAUDA \citep{cui2020gradually}      	&161.7$\pm$0.1& {0.8813}$\pm$0.0012&	 {33.27}$\pm$0.06 & 10.62$\pm$0.13\\ 

JGLA \citep{yarram2022joint}  	&160.4$\pm$0.4 & {0.8815}$\pm$0.0012&	 {33.24}$\pm$0.05 & 10.67$\pm$0.10   \\\hline

AC:GST  &\textbf{157.1}$\pm$0.3&	\textbf{0.9217}$\pm$0.0012& \textbf{35.92}$\pm$0.05& \textbf{13.70}$\pm$0.12\\
AC:GST-$\mathcal{A}_{\bm{\theta}}$      & {157.8}$\pm$0.2&	 {0.9151}$\pm$0.0015&  {35.40}$\pm$0.07&  {13.19}$\pm$0.13\\
AC:GST-C    & {158.0}$\pm$0.2&	 {0.9183}$\pm$0.0010&  {35.65}$\pm$0.03 &  {13.28}$\pm$0.11\\ 

BM:GST \citep{liu2021generative}   & {158.6}$\pm$0.2&	 {0.9078}$\pm$0.0011&  {34.48}$\pm$0.05  & 12.63$\pm$0.12\\
BM:GST-A      & {159.5}$\pm$0.3&	 {0.8997}$\pm$0.0011&  {34.03}$\pm$0.04  & 12.03$\pm$0.12\\
BM:GST-E    & {159.8}$\pm$0.1&	 {0.9026}$\pm$0.0013&  {34.05}$\pm$0.05  & 11.95$\pm$0.11\\\hline

{Target Supervised}  & {148.2}$\pm$0.2&	 {0.9516}$\pm$0.0012&  {37.42}$\pm$0.05  & 15.37$\pm$0.13\\\hline

\end{tabular} 
}
\label{tabel:1} 
\end{table}

As the source domain, we acquired a total of 1,768 paired tagged and cine tongue MR slices from a cohort of 10 healthy subjects at UMB. The data was acquired using a segmented gradient-echo sequence on a Siemens 3.0T TIM Trio system with a 12-channel head coil and a 4-channel neck coil~\citep{xing2016analysis}. The imaging parameters were as follows: a field of view of 240$\times$240 mm, an in-plane resolution of 1.88$\times$1.88 mm, and a slice thickness of 6 mm. The image sequence was obtained at a rate of 26 fps, synchronized to the subject’s verbal response. Both cine and tagged MR images were acquired in the same spatiotemporal coordinate space.

For the target domain, i.e., cross-scanner target domain 1, we collected a total of 1,014 paired tagged and cine tongue MR slices from a cohort of 5 different healthy subjects at UMB. We used a Siemens 3.0T Prisma scanner with a 64-channel head and neck coil, and acquired the images using the same sequence as in the source domain~\citep{xing2016analysis}. Due to the use of different scanners, there were appearance discrepancies between the datasets acquired from the two scanners. The imaging parameters were as follows: a field of view of 240$\times$240 mm, an in-plane resolution of 1.88$\times$1.88 mm, and a slice thickness of 6 mm. The image sequence was obtained at the rate of 26 fps, synchronized with the subject’s verbal response. Both cine and tagged MR images were in the same spatiotemporal coordinate space. We performed five-fold cross-validation on the 5 subjects in the target domain. {In each round, we used one subject for testing, while the other four were used for training and validation. Notably, we followed the test-time adaptation protocol, where the source domain could be considered as an auxiliary set with independent subjects for target training/validation/testing.}

In Fig.~\ref{fig:results1}, we show a qualitative comparison of the synthesis results using the proposed AC:GST and the other comparison methods, including source domain Pix2Pix~\citep{isola2017image} without UDA training, vanilla adversarial UDA (ADDA)~\citep{tzeng2017adversarial}, and gradually adversarial UDA (GAUDA)~\citep{cui2020gradually}. It is worth noting that the domain-wise distribution alignment loss in adversarial UDA methods often leads to the presence of hallucinated content, resulting in significant differences in the shape and texture of the tongue between the real cine MR images. In contrast, the proposed GST UDA does not rely on adversarial training, leading to the synthesis of visually pleasing slices with more consistent anatomical structure, as demonstrated in Fig.~\ref{fig:results1}. Such consistent anatomical structure is essential for subsequent analyses~\citep{xing2016analysis}.

\begin{figure}[t]
\begin{center}
\includegraphics[width=1\linewidth]{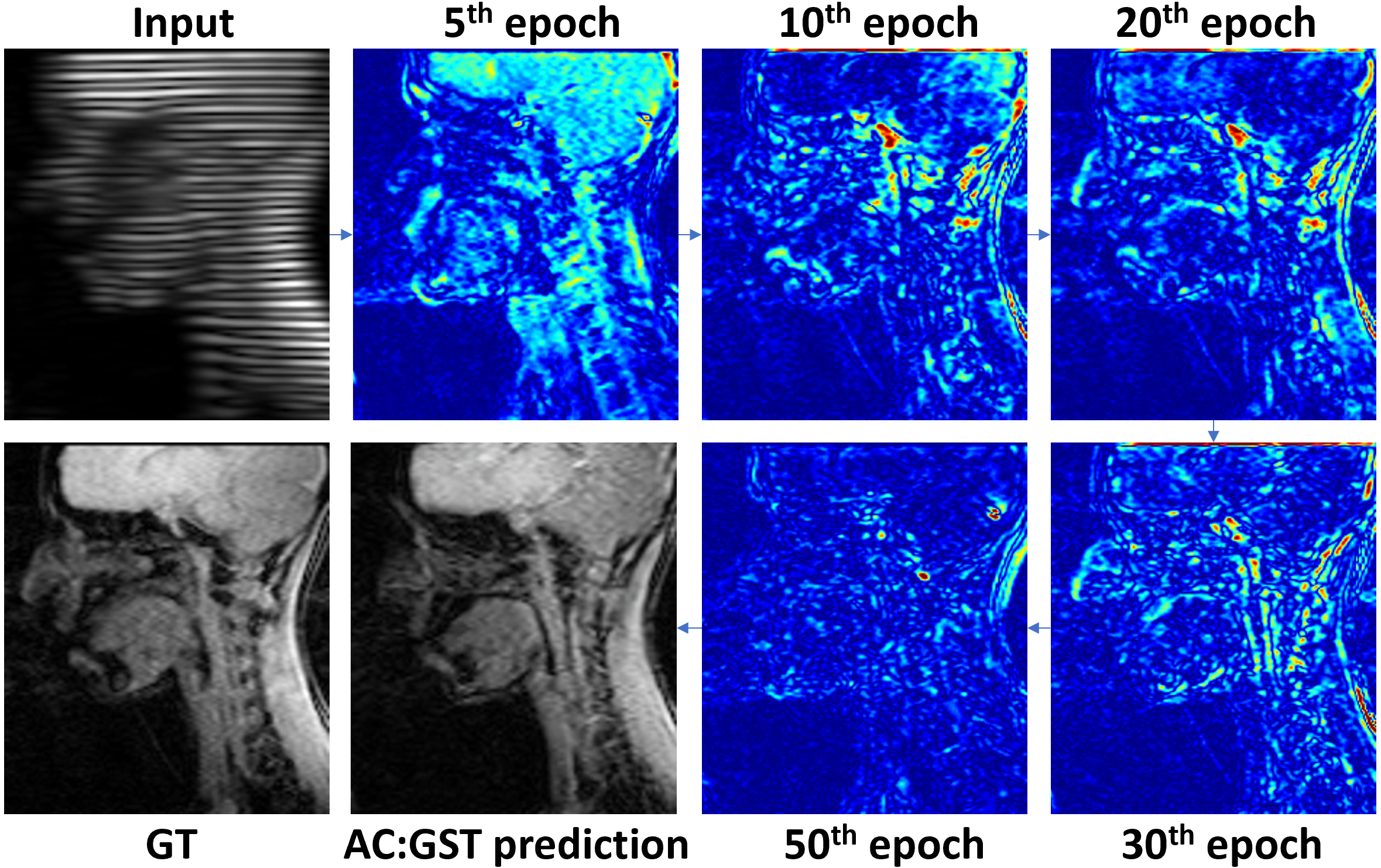}
\end{center}\vspace{-10pt}
\caption{{Qualitative example of uncertainty measurement during training epochs for the UMB cross-scanner tagged-to-cine MR translation task.}} 
\label{fig:visual}
\end{figure}

\begin{figure*}[t]
\begin{center}
\includegraphics[width=1\linewidth]{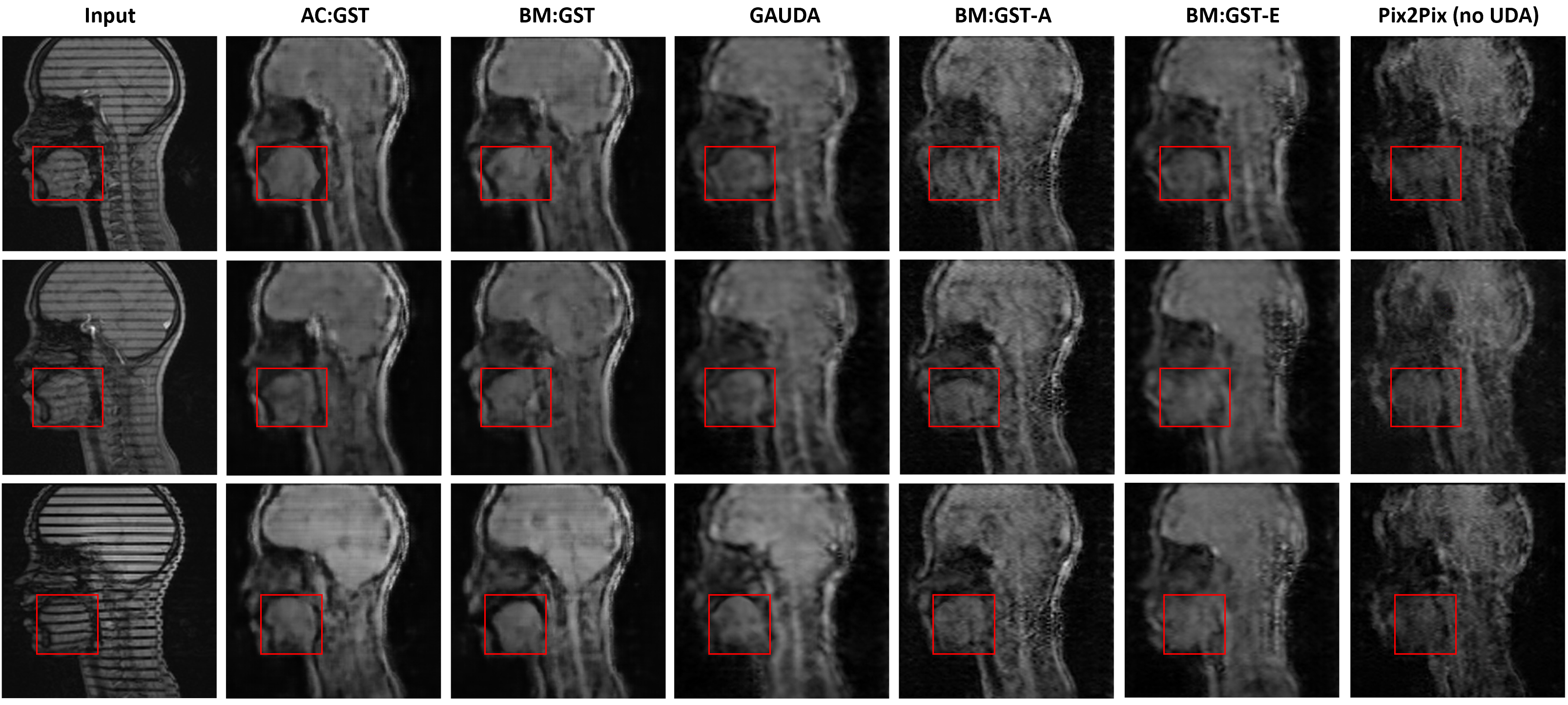} 
\end{center}  
\caption{Comparison of various UDA methods on the UMB-MGH cross-center tagged-to-cine MR image translation task, including our proposed AC:GST, BM:GST, BM:GST-A, and BM:GST-E, as well as adversarial UDA~\citep{cui2020gradually} and Pix2Pix~\citep{isola2017image} without adaptation. The red rectangle indicates the tongue region.}  
\label{fig:results3}
\end{figure*} 

In the ablation study, the results in Fig.~\ref{fig:results2} demonstrate that AC:GST with the self-attention continuous reliability mask outperforms AC:GST-$\mathcal{A}_{\bm{\theta}}$, AC:GST-C, and BM:GST. Additionally, the comparison between BM:GST-E, BM:GST-A, and BM:GST demonstrates the advantage of considering both aleatoric and epistemic uncertainties for the mask.

We expect that the synthesized images would have realistic and structurally consistent textures as compared with their corresponding ground truth images. To quantitatively assess our framework, we adopted well-known evaluation metrics including structural similarity index measure (SSIM), peak signal-to-noise ratio (PSNR), mean L1 error, and inception score (IS)~\citep{liu2021dual}. Table \ref{tabel:1} lists numerical comparisons using a total of five testing subjects in the target domain. When compared with~\cite{he2021autoencoder}, UDA methods that leveraged source domain data during the adaptation stage~\citep{tzeng2017adversarial,cui2020gradually,yarram2022joint} achieved marked improvements in the measures listed above. Our GST surpassed GAUDA~\citep{cui2020gradually}, and ADDA~\citep{tzeng2017adversarial} in terms of SSIM, PSNR, L1 error, and IS by a large margin. {The averaged results of the target domain supervised model are also provided, which can be regarded as an ``upper-bound".} {To show statistical significance, we performed a one-tailed paired t-test for each of the L1, SSIM, PSNR, and IS metrics, comparing improvements over BM:GST with AC:GST, yielding p-values of $0.026$, $0.0085$, $0.0042$, and $0.0014$, respectively. When comparing AC:GST with JGLA, we obtained p-values of L1, SSIM, PSNR, and IS metrics of $0.013$, $0.0037$, $2.63\times10^{-4}$, and $4.70\times10^{-6}$, respectively.}

We empirically set weights $\beta=1$ and $K=20$ based on validation to balance the absolute value of each loss. While a larger $K$ can help approximate the true uncertainty better, the computing cost increases linearly with $K$. Therefore, it is necessary to find a reasonable value of $K$ that balances between uncertainty estimation and computation efficiency.

The multiplier $\beta\in\mathbb{R}^+$ was used to balance between $\frac{1}{\sigma^2_{t,n}}||(\hat{y}_{t,n}-\tilde{y}_{t,n})m_{t,n}||^2_2$ and $\text{log} \sigma^2_{t,n}$. Typically, networks are not sensitive to the Lagrangian multiplier within a large range. To investigate the effect of $\beta$ on the performance of AC:GST in the cross-scanner tagged-to-cine UDA task, we carried out a sensitivity analysis, as shown in Fig. \ref{sensi1}. The results demonstrate that the performance of AC:GST is stable for $\beta\in[1,1.5]$. The number of MC dropouts can affect the epistemic uncertainty estimation. Larger values of $K$ can help in approximating the true uncertainty, but it will lead to an increase in computation cost linearly. Therefore, it is necessary to find a suitable value for $K$ to balance uncertainty estimation and computation efficiency. In Table \ref{sensi1}, we provide a detailed sensitivity analysis of $K$ for AC:GST in the cross-scanner tagged-to-cine UDA task. It can be observed that the performance is relatively stable for $K\geq20$. In BM:GST, the meta portion parameter $\rho$ is gradually increased from 30\% to 80\% over the training to select more pseudo-labels, as in \citep{zou2019confidence,liu2020energy}. However, in AC:GST, we do not rely on $\rho$ for binary selection.

{As shown in Fig. \ref{fig:visual}, the uncertainty gradually decreases as the training progresses. In the early stages, e.g., the 5th epoch, our uncertainty map focuses on the ROI, after which it concentrates on refining the shape boundary. Of note, the model is also constrained by attention maps. In our tongue and brain MR image translation tasks, the ROI is the anatomical part, while the remaining blank region should follow an easy identical mapping. Therefore, even in the initial epochs, the uncertainty will focus on the ROI region and will not be high in everywhere. For the more fine-grained uncertainty control, our method can also avoid the impact of large uncertainty in a wide region. Specifically, the binary mask uses $\rho$ to select the top-ranked confident pixel. Therefore, our mask is only related to the relative uncertainty with the ranking protocol, not the absolute uncertainty value. For the attentive continuous mask, we have another attention module to constrain the model to focus on ROI, and the e-index function in Eq.~(10) can also eliminate small uncertainty values. 
}
%In Table \ref{tabel:71}, we provide a detailed sensitivity analysis of $\rho$ for BM:GST in the cross-scanner tagged-to-cine UDA task. 

\begin{figure*}[t]
\begin{center}
\includegraphics[width=1\linewidth]{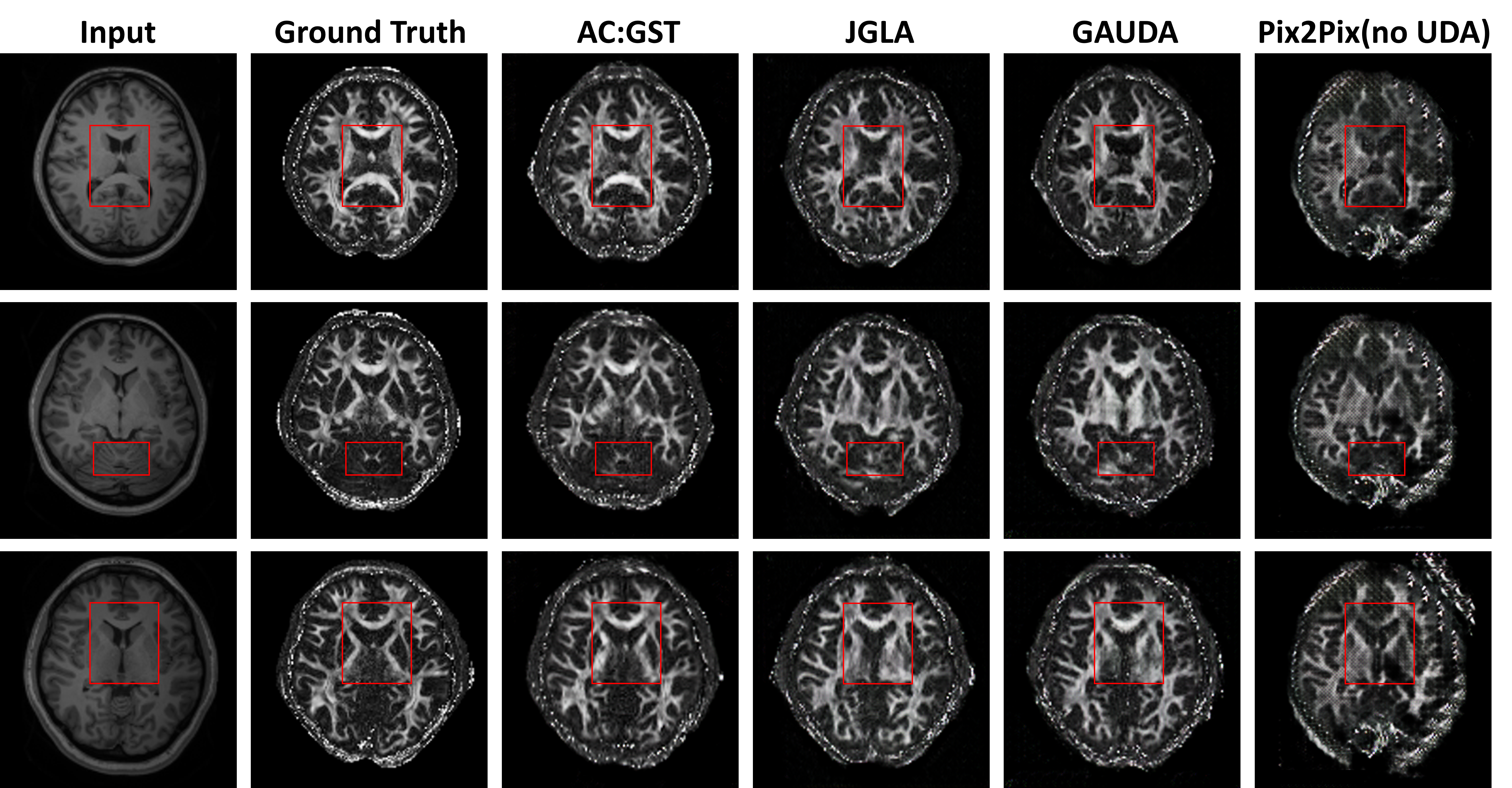} 
\end{center}  
\caption{Comparison of our framework with the adversarial UDA~\citep{cui2020gradually,yarram2022joint} and Pix2Pix~\citep{isola2017image} without adaptation for the WUMinn-MGH cross-center T1-to-FA brain MR image translation. The red rectangle highlights the region with noticeable artifacts.}  
\label{fig:results4}
\end{figure*}

%\begin{table}[t]
%\centering
%\caption{Numerical comparisons of the cross-scanner and cross-center UDA with or without additional cross-domain discriminator for adversarial UDA.}  \vspace{+3pt}
%\resizebox{1\linewidth}{!}{
%\begin{tabular}{c|c|c|c|c||ccc|c|cc}
%\hline
% &\multicolumn{4}{c||}{Cross-scanner} & Cross-center\\ \hline
%Methods& L1~$\downarrow$& SSIM~$\uparrow$ &	PSNR~$\uparrow$ &  IS~$\uparrow$&  IS~$\uparrow$\\\hline\hline
%BM:GST~\citep{liu2021generative}   & {158.6}$\pm$0.2&	 {0.9078}$\pm$0.0011&  {34.48}$\pm$0.05&   {12.63}$\pm$0.12&  {9.76}$\pm$0.11\\
%BM:GST+Dis  & {157.9}$\pm$0.1&	 {0.9080}$\pm$0.0012&  {34.47}$\pm$0.07&   {12.64}$\pm$0.13& {9.76}$\pm$0.12\\\hline
%AC:GST  & {157.1}$\pm$0.3&	 {0.9217}$\pm$0.0012&  {35.92}$\pm$0.05& {13.70}$\pm$0.12 &  {9.86$\pm$0.13}\\ 
%AC:GST+Dis  & {157.2}$\pm$0.2& {0.9219}$\pm$0.0015& {35.92}$\pm$0.06&  {13.71}$\pm$0.10 &  {9.85$\pm$0.11}\\\hline
%\end{tabular}
%}
%\label{tabel:42}
%\end{table}

%In addition, we do not observe significant improvements of adding additional adversarial UDA over our BM:GST and AC:GST. The quantitative results are shown in Table. \ref{tabel:42}. GST+Dis indicates using source domain only discriminator as in Pix2Pix. The improvement was not significant, while the additional adversarial objective required 50 min for GST to converge. Therefore, we used Pix2Pix as a good initialization and dropped its source domain only discriminator at the UDA stage. 

\subsection{UMB-MGH tagged-to-cine MR translation}

 To further demonstrate the generality of our framework in handling larger domain gaps, we collected 120 tagged 2D MR slices from one subject at Massachusetts General Hospital (MGH) as the cross-center target domain. The MRI scanning was performed using a Siemens Skyra 3T scanner with a 64-channel head and neck coil, employing a gradient recalled echo sequence. The MRI acquisition parameters included a 128$\times$128 field of view, 12 slices of 6 mm thickness acquired at 10 equally spaced time points, 86 ms apart, synchronized to the subject's verbal response. The in-plane resolution was 2 mm$\times$2 mm, and the tag spacing was 12 mm. Due to differences in soft tissue contrast, as well as imaging parameters such as tag spacing and field of view, there were notable discrepancies between the data collected at MGH and UMB. As there were no paired cine MR images available as ground truth in the target domain, we set $\beta=1$ and $K=20$, following the settings used for the cross-scanner translation task.

\begin{table}[t]
\centering
\caption{Numerical comparisons and ablation studies of the UMB-MGH cross-center tongue tagged-to-cine MR translation task} \vspace{+5pt}
\resizebox{1\linewidth}{!}{
\begin{tabular}{l|ccl|c}
\cline{1-2}\cline{4-5}
Compared Methods& IS~$\uparrow$ & &Proposed Methods& IS~$\uparrow$    \\\cline{1-2}\cline{4-5}

w/o UDA~\citep{isola2017image} &5.32$\pm$0.11 &&AC:GST  & \textbf{9.86$\pm$0.13}\\\cline{1-2}

ASS~\citep{he2021autoencoder} &6.94$\pm$0.12 &&AC:GST-$\mathcal{A}_{\bm{\theta}}$&9.82$\pm$0.12\\\cline{1-2}

ADDA~\citep{tzeng2017adversarial} &8.69$\pm$0.10 &&BM:GST~\citep{liu2021generative}  &9.76$\pm$0.11\\ 

GAUDA~\citep{cui2020gradually}  &8.83$\pm$0.14  &&BM:GST-E &9.54$\pm$0.13\\

JGLA~\citep{yarram2022joint} &8.75$\pm$0.12 &&BM:GST-A &9.58$\pm$0.12\\\cline{1-2}\cline{4-5}
 
\end{tabular}
}
\label{tabel:2} 
\end{table}

In Fig.~\ref{fig:results3}, we provide quantification results of aleatoric and epistemic uncertainties alongside qualitative comparisons between GAUDA and the ablation study of the self-attention continuous reliability mask. We show that the anatomic structure of the tongue is better synthesized using the AC:GST framework, compared with the other comparison methods. Because of the large domain gaps in the datasets between the two sites, the image quality of the UMB-MGH cross-center translation setting was inferior to that of the within UMB cross-scanner translation setting, as visually assessed. 

Of note, in the MGH dataset, we do not have paired ground truth data. Therefore, we provide quantitative comparisons using the inception score (IS), which does not require paired labels \citep{liu2021dual}. As shown in Table \ref{tabel:2}, our proposed GST outperforms the adversarial UDA approaches, i.e., GAUDA and ADDA \citep{cui2020gradually,tzeng2017adversarial}, in a consistent manner, indicating the effectiveness of self-training for this generative image translation UDA task. % in addition to discriminative self-training methods~\citep{zou2019confidence,liu2020energy}.

\subsection{WUMinn-MGH cross-center T1-to-FA MR translation}

\begin{figure*}[t]
\begin{center}
\includegraphics[width=1\linewidth]{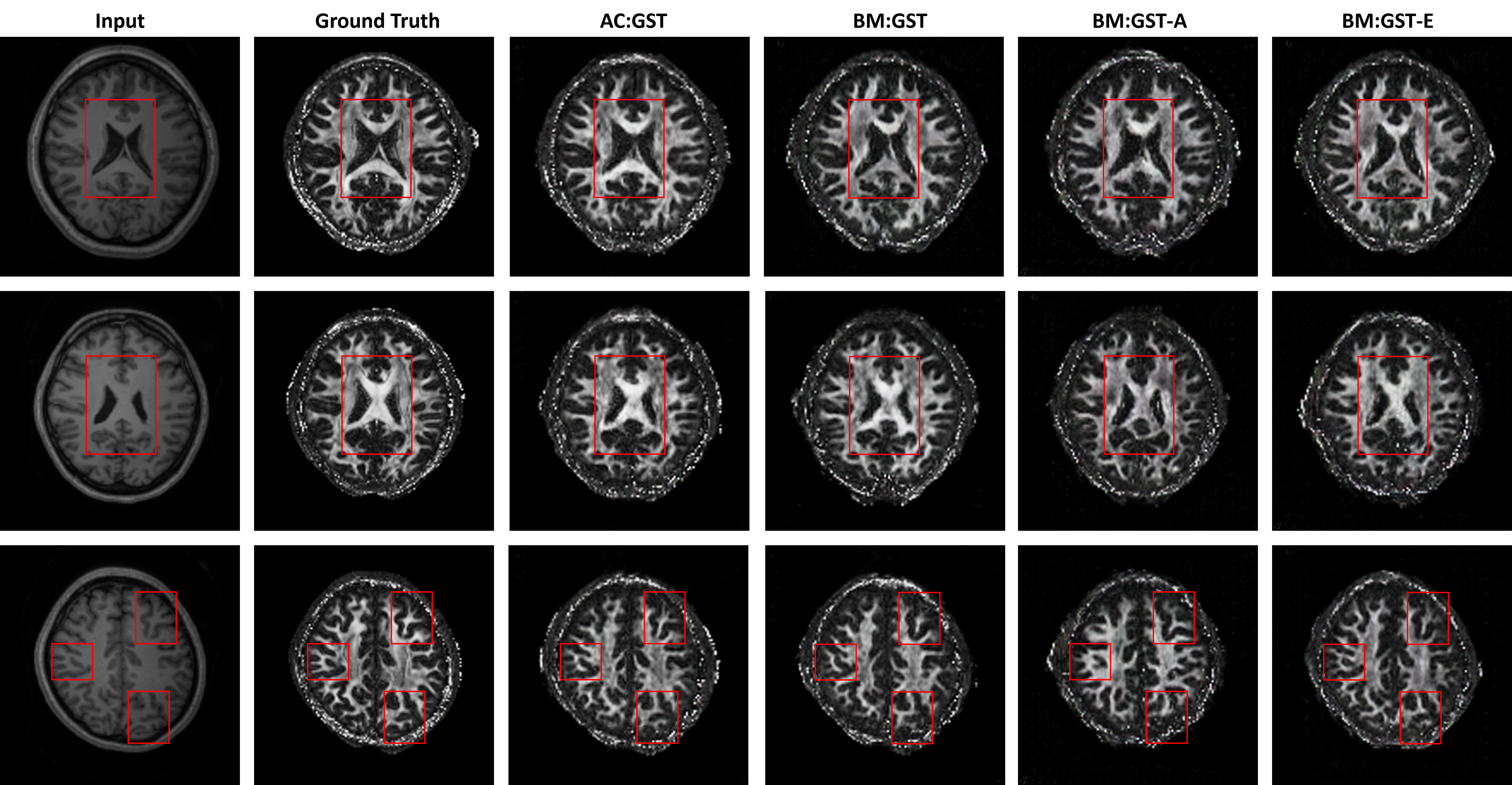} 
\end{center}  
\caption{Ablation studies of our proposed AC:GST, BM:GST, BM:GST-A, and BM:GST-E for WUMinn-MGH cross-center T1-to-FA brain MR image translation. The red rectangle highlights the region with noticeable artifacts.}   
\label{fig:results5}
\end{figure*} 

Diffusion MRI is a non-invasive imaging method that enables the measurement of water molecule diffusion in tissue, allowing for the quantification of microstructural tissue properties. Scalar measures of diffusion computed from diffusion MRI, such as fractional anisotropy (FA) and mean diffusivity, have become widely used. However, collecting high-quality diffusion MRI data can be challenging and time-consuming. Moreover, diffusion MRI can be affected by various artifacts, such as motion and susceptibility artifacts, which can hinder subsequent analyses~\citep{gu2019generating}. Therefore, synthesizing scalar diffusion quantities from structural MR scans, which are easier to acquire than diffusion MRI, would be useful~\citep{gu2019generating}.

\begin{table}[t]
\centering
\caption{Numerical comparisons and ablation studies of the WUMinn-MGH cross-center brain T1-to-FA MR translation task} \vspace{+5pt}
\resizebox{1\linewidth}{!}{
\begin{tabular}{l|c|c|c|c}
\hline

Methods& L1~$\downarrow$& SSIM~$\uparrow$ &	PSNR~$\uparrow$ &  IS~$\uparrow$\\\hline\hline
w/o UDA~\citep{isola2017image}     	& {158.5}$\pm$0.4&	0.8264$\pm$0.0017&	25.49$\pm$0.06&	9.27$\pm$0.14  \\\hline

ASS~\citep{he2021autoencoder} & {132.8}$\pm$0.3&	0.8847$\pm$0.0014& 31.63$\pm$0.04&	11.54$\pm$0.10  \\\hline

ADDA~\citep{tzeng2017adversarial}	&129.2$\pm$0.4&	0.9131$\pm$0.0016&	34.85$\pm$0.05&	13.28$\pm$0.12\\
GAUDA~\citep{cui2020gradually}      	&129.0$\pm$0.5& {0.9174}$\pm$0.0015&	 {35.24}$\pm$0.06 & 13.46$\pm$0.13\\ 

JGLA~\citep{yarram2022joint}  	&128.6$\pm$0.2 & {0.9205}$\pm$0.0016&	 {35.22}$\pm$0.03 & 13.50$\pm$0.12  \\\hline

AC:GST  &\textbf{121.2}$\pm$0.3&	\textbf{0.9692}$\pm$0.0014& \textbf{38.65}$\pm$0.04& \textbf{16.42}$\pm$0.14\\
AC:GST-$\mathcal{A}_{\bm{\theta}}$      & {122.0}$\pm$0.3&	 {0.9637}$\pm$0.0014&  {37.94}$\pm$0.04&  {16.05}$\pm$0.12\\
AC:GST-C    & {121.7}$\pm$0.3&	 {0.9641}$\pm$0.0014&  {38.26}$\pm$0.06 &  {15.97}$\pm$0.13\\ 

BM:GST~\citep{liu2021generative}   & {122.8}$\pm$0.3&	 {0.9563}$\pm$0.0014&  {37.65}$\pm$0.04  & 15.12$\pm$0.12\\
BM:GST-A      & {123.5}$\pm$0.4&	 {0.9488}$\pm$0.0016&  {37.30}$\pm$0.06  & 14.61$\pm$0.11\\
BM:GST-E    & {122.9}$\pm$0.2&	 {0.9532}$\pm$0.0015&  {37.16}$\pm$0.04  & 14.69$\pm$0.13\\\hline

{Target Supervised}  & {114.6}$\pm$0.3&	 {0.9815}$\pm$0.0013&  {42.27}$\pm$0.06  & 18.83$\pm$0.10\\\hline

\end{tabular}
}
\label{tabel:3} 
\end{table}

\begin{figure}[t]
\begin{center}  \vspace{+3pt}
\includegraphics[width=1\linewidth]{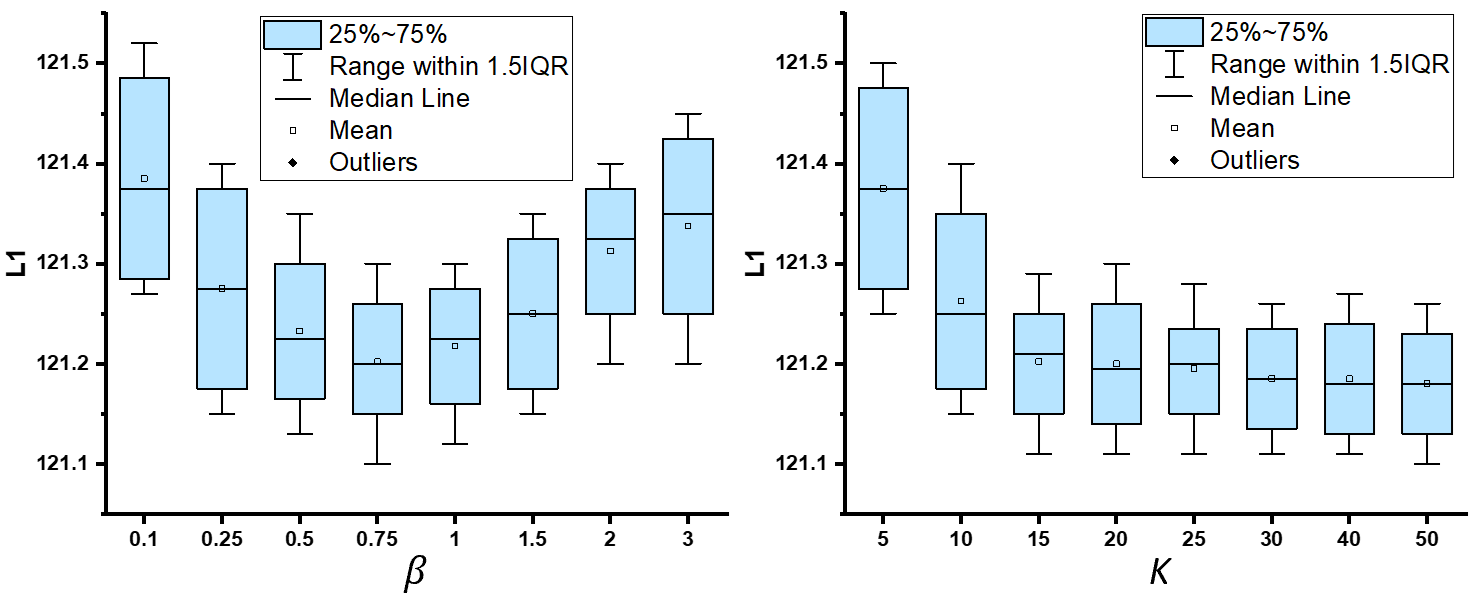}
\end{center}  \vspace{-10pt}
\caption{Sensitivity analysis of $\beta$ and $K$ in AC:GST with respect to L1 error for WUMinn-MGH cross-center T1-to-FA MR image translation.}  
\label{sensi2}\end{figure}

As the source domain data, we used data from a total of five adolescents (14--15 years old) from the WUMinn HCP database\footnote{\href{https://www.humanconnectome.org/study-hcp-lifespan-pilot}{Link to WUMinn HCP database}}, which was acquired using a 3T scanner with a maximum gradient strength of 70--100mT/m for diffusion MRI. As the target domain, we used data from a total of ten adult subjects (20--59 years old) from the MGH HCP database\footnote{\href{https://www.humanconnectome.org/study/hcp-young-adult/document/mgh-adult-diffusion-data-acquisition-details}{Link to MGH HCP database}} acquired using the customized MGH Siemens 3T Connectome scanner, which has a maximum gradient strength of 300mT/m. Prior to synthesis, each T1-weighted MRI was affinely registered to its corresponding b0 image using the cross-correlation similarity metric with the ANTs toolkit~\citep{avants2011reproducible}. Each subject in the WUMinn and MGH HCP databases has a total of 93 and 96 paired slices, respectively. The diffusion tensor was reconstructed using the DIPY library~\citep{garyfallidis2014dipy} with the tensor model~\citep{basser1994mr}, and then FA volumes were computed. {For evaluation, we adopted a five-fold cross-validation strategy for the ten subjects in the target domain, using two subjects for testing in each round, while using the remaining eight subjects for training and validation.}

We provide qualitative comparisons against the adversarial UDA methods GAUDA~\citep{cui2020gradually} and ADDA~\citep{tzeng2017adversarial} in Fig.~\ref{fig:results4}. {We carried out one-tailed paired t-tests for each of the L1, SSIM, PSNR, and IS metrics to compare improvements over BM:GST with AC:GST. The resulting p-values were $0.014$, $0.012$, $0.0076$, and $6.09\times10^{-4}$, respectively, indicating statistical significance. When comparing AC:GST with JGLA, we obtained p-values of $0.0037$, $0.0016$, $2.65\times10^{-4}$, and $5.49\times10^{-5}$ for the L1, SSIM, PSNR, and IS metrics, respectively, also indicating statistical significance.} The ablation study results are shown in Fig.~\ref{fig:results5}. In addition, the corresponding quantitative results are given in Table \ref{tabel:3}. In contrast to the tagged-to-cine MR translation task, our AC:GST framework outperformed the Pix2Pix baseline and adversarial UDA methods by a large margin. Of note, we empirically set weight $\beta=0.75$ and $K=15$, and provide the sensitivity analysis in Fig. \ref{sensi2}. The results indicate that the performance of our model is relatively consistent for $\beta\in[0.5,1.0]$ and $K\geq15$.

\section{Discussion and Conclusion} 

In this work, we proposed a novel generative self-training framework with adaptively learned attention for two unsupervised domain adaptive medical image translation tasks. By equipping the self-training framework with a unified uncertainty quantification scheme for both epistemic and aleatoric uncertainties in a UDA setting, we were able to adaptively control the generative pseudo-label supervision according to the reliability of the pseudo-labels. This was the first attempt at achieving the UDA of tagged-to-cine and structure-to-diffusion (e.g., T1-to-FA) image translation. Our experimental results clearly showed that, when quantitatively and qualitatively assessed, our framework yielded superior translation performance over popular adversarial UDA methods. The synthesized cine MR or FA images with test time UDA could be utilized to delineate the tongue and to observe surface motion or quantify micro-structural tissue properties, without the need for additional acquisition costs and time~\citep{gu2019generating}.

This work aimed to achieve adaptation with limited unpaired target data and fast convergence in the test-time UDA setting~\citep{karani2021test}. We were able to improve synthesis performance for an unseen subject, by retraining with its input slices. Many UDA methods use both paired source domain data and unpaired target domain data for joint training at the adaptation stage. In comparison, our AC:GST and BM:GST training took approximately 50 and 30 minutes, respectively, for within UMB cross-scanner tagged-to-cine UDA, whereas GAUDA~\citep{cui2020gradually} took 2 hours to train. Furthermore, synthesizing one slice in testing took only about 0.1 seconds. We note that~\cite{zuo2021information} is not applicable for test-time UDA evaluation since it requires target domain sample labels. Additionally, the performance of source-free test-time adaptation~\citep{he2021autoencoder,liu2021Off-the-Shelf,karani2021test} may degrade without the supervision of source domain data in adaptation.

Our framework has notable features as follows. First, instead of setting pseudo-labels as a variable with one-hot or all zero vectors as in conventional self-training methods~\citep{zou2019confidence,wei2021theoretical,liu2020energy}, our GST framework leverages the reliability mask as a learned variable to help control pseudo-label selection. In this way, we can flexibly adjust the contribution of each pseudo-labeled pixel for the image translation tasks. Second, we incorporate both aleatoric and epistemic uncertainties in our generative self-training UDA using a Bayesian reliability mask in a principled manner. Specifically, both types of uncertainty are utilized to compensate for the inherently noisy pseudo-labels during self-training. The results of our ablation studies in Tables~\ref{tabel:1}-\ref{tabel:3} show that taking both of them into consideration can contribute to improved performance. Figures~\ref{fig:results2}, \ref{fig:results3}, and \ref{fig:results5} illustrate that when aleatoric uncertainty resulting from inaccurate labels is not accounted for, BM:GST-A produces results with slight distortions in shape and boundary. On the other hand, when epistemic uncertainty is not considered, BM:GST-E tends to generate noisier outputs compared with BM:GST. An advanced uncertainty measurement can potentially be added to achieve either more accurate quantification or faster inference. A third notable feature in our GST framework is that the correlation between the quantified uncertainty of pseudo-labels and the reliability mask is flexibly defined. Although the binary reliability mask with threshold-based pseudo-label selection in Eq.~(\ref{bm}) can be a straightforward solution to convert the quantified uncertainty as a reliability measurement of pseudo-labels, it can be too rough to be applied to our well-quantified uncertainty. The fine-grained continuous reliability mask that we propose contributes to the better performance of AC:GST-$\mathcal{A}_{\bm{\theta}}$ over BM:GST, as shown in Tables~\ref{tabel:1}-\ref{tabel:3}.  A fourth notable feature in our framework is that self-attention was added to adaptively emphasize the regions of interest. Medical imaging data, particularly MR imaging data, often have homogeneous background areas. As a result, in synthesis tasks, pixels in the background region usually undergo a simple identity mapping, which is easy to learn and has low uncertainty, resulting in high reliability predictions in the early epochs. This dominance of easily learned pixels could result in their overrepresentation in the GST, without compensation. Our results, as shown in Tables~\ref{tabel:1}-\ref{tabel:3}, consistently demonstrated that our AC:GST, which includes an adaptive emphasis on regions of interest, outperformed AC:GST-$\mathcal{A}_{\bm{\theta}}$.
 
A few important components have not been fully investigated in the present work. First, although we applied our framework to MR image translation tasks, the idea of the variable reliability mask and the continuous reliability mask may also be helpful for discriminative self-training UDA with more accurate uncertainty measurement than the maximum softmax probability. Second, it is worth noting that our investigation only focused on the scenario, where the source and target domain models share the same backbone, which is a common approach in UDA. However, different backbones in the source and target domain can also be used, by initializing the target model with knowledge distillation~\citep{liu2022unsupervised}. Third, since our framework has shown superior performance for the challenging test-time UDA task, our framework is likely to yield better performance, when taking more unpaired target domain subjects in training. We will investigate different input data and UDA scenarios to confirm the full potential and utility of the proposed framework in our future work.  %Fourth, our framework was implemented and tested in 2D with MRI datasets. Extension to a 3D translator with different imaging modalities could be challenging, as it requires complicated network structures alongside larger datasets and computational resources. This is subject to our future work.

\section*{Acknowledgments}

We gratefully acknowledge funding support from NIH R01DC014717, R01DC018511, R01CA133015, and P41EB022544.

\bibliographystyle{model2-names.bst}\biboptions{authoryear}
\bibliography{refs}

\end{document}